\def\draftversion{false}
\RequirePackage{ifthen}
\ifthenelse{\equal{\draftversion}{true}}{
  \documentclass[aps,prx,10pt,galley,amsmath,amssymb,
                 superscriptaddress]{revtex4}
}{
  \documentclass[aps,prx,10pt,twocolumn,amsmath,amssymb,
                 longbibliography,superscriptaddress]{revtex4-1}
}

\usepackage{graphicx}% Include figure files
\usepackage[usenames,dvipsnames]{color} % colors
\usepackage{bm} % bold math
\usepackage{hyperref} % Include hyperlinks
\hypersetup{		  % Make hyperlinks colorful
    colorlinks=true,
    linkcolor=blue,
    filecolor=blue,      
    urlcolor=blue,
    citecolor=blue
}
\usepackage{array}
\usepackage{soul}

\ifthenelse{\equal{\draftversion}{true}}{
  \marginparwidth 2.7in
  \marginparsep 0.5in
  \newcounter{comm} % counter for commentaries
  \def\commnext{\stepcounter{comm}}
  \def\commtext{{\bf\color{blue}[\arabic{comm}]}}
  \def\commmar{{\bf\color{blue}[\arabic{comm}]}}
  \def\dvm#1{\commnext\marginpar{\small DV\commmar: #1}\commtext}
  \def\nvm#1{\commnext\marginpar{\small NV\commmar: #1}\commtext}
  \def\ism#1{\commnext\marginpar{\small IS\commmar: #1}\commtext}
  \def\mlab#1{\marginpar{\small\bf #1}}
  
  \def\parsedate #1:20#2#3#4#5#6#7#8\empty{#4#5/#6#7/20#2#3}
  \def\moddate{\expandafter\parsedate\pdffilemoddate{\jobname.tex}\empty}

  \newcommand{\seclabel}[1]{\label{sec:#1}\Red{\small\;\;[Sec:~#1]}}
}{
  \def\dvm#1{}
  \def\nvm#1{}
  \def\ism#1{}
  \def\mlab#1{}
  
  \newcommand{\seclabel}[1]{\label{sec:#1}}
}

\def\Red#1{\textcolor{red}{#1}}

\newcommand{\sref}[1]{Sec.~\ref{sec:#1}}

\newcommand{\fref}[1]{Fig.~\ref{fig:#1}}

\newcommand{\beq}{\begin{equation}}
\newcommand{\eeq}{\end{equation}}
\newcommand{\bea}{\begin{eqnarray}}
\newcommand{\eea}{\end{eqnarray}}

\newcommand{\ignore}[1]{}
\def\zt{\mathbb{Z}_2}

\def\k{{\bf k}}

\def\ket#1{\vert#1\rangle}

\begin{document}

%===========================%
% TITLE PAGE                %
%===========================%

\title{Engineering magnetic topological insulators in Eu$_5$\textit{M$_2$X$_6$} Zintls}

\author{Nicodemos Varnava}
\thanks{Corresponding author} 
\email{nvarnava@physics.rutgers.edu}
\affiliation{
Department of Physics \& Astronomy, Rutgers University,
Piscataway, New Jersey 08854, USA}

\author{Tanya Berry}
\affiliation{Institute for Quantum Matter and William H. Miller III Department of Physics and Astronomy, The Johns Hopkins University, Baltimore, Maryland 21218, USA}
\affiliation{Department of Chemistry, The Johns Hopkins University, Baltimore, Maryland 21218, USA}
\affiliation{Department of Materials Science and Engineering, The Johns Hopkins University, Baltimore, Maryland 21218, United States
}

\author{Tyrel M. McQueen}
\affiliation{Institute for Quantum Matter and William H. Miller III Department of Physics and Astronomy, The Johns Hopkins University, Baltimore, Maryland 21218, USA}
\affiliation{Department of Chemistry, The Johns Hopkins University, Baltimore, Maryland 21218, USA}
\affiliation{Department of Materials Science and Engineering, The Johns Hopkins University, Baltimore, Maryland 21218, United States
}

\author{David Vanderbilt}
\affiliation{
Department of Physics \& Astronomy, Rutgers University,
Piscataway, New Jersey 08854, USA}

\begin{abstract}

Magnetic topological insulator provide a prominent material platform for quantum anomalous Hall physics and axion electrodynamics. However, the lack of material realizations with cleanly gapped surfaces hinders technological utilization of these exotic quantum phenomena. Here, using the Zintl concept and the properties of non-symmorphic space groups, we computationally engineer magnetic topological insulators. Specifically, we explore Eu$_5M_2X_6$ (\textit{M}=metal, \textit{X}=pnictide) Zintl compounds and find that Eu$_5$Ga$_2$Sb$_6$, Eu$_5$Tl$_2$Sb$_6$ and Eu$_5$In$_2$Bi$_6$ form stable structures with non-trivial $\zt$ indices.  We also show that epitaxial and uniaxial strain can be used to control the $\zt$ index and the bulk energy gap. Finally, we discuss experimental progress towards the synthesis of the proposed candidates and provide insights that can be used in the search for robust magnetic topological insulators in Zintl compounds.

\end{abstract}

\maketitle

%===========================%
% MAIN TEXT                 %
%===========================%

%=================================================
\section{Introduction}\seclabel{intro}
%=================================================

Topology and symmetry have played a profound role in shaping modern condensed matter physics and materials science. One of the most paradigmatic examples are the 3D $\zt$ topological insulators (TIs) protected by time-reversal symmetry\cite{hasan-rmp10,qi-prm11}. Owing to their bulk band topology, these insulators possess an odd number of massless Dirac fermions with spin–momentum locking at their surfaces. 

Although 3D topological insulators were originally proposed in time-reversal invariant systems, the $\zt$ index can also be protected by a symmetry other than simple time-reversal (TR)\cite{qi-prb08,essin-prl09+e}. These symmetries comprise of the proper rotations composed with TR and the
improper rotations not composed with TR\cite{varnava-prb20}. If one or more of these
symmetries is present in the magnetic point group, then the $\zt$ index is well defined, and if this index is nontrivial we refer to it as a ``magnetic TI" or equivalently an ``axion insulator". Gapped surfaces of magnetic TIs can appear, in which case they will exhibit half-integer surface anomalous Hall conductivity (AHC), whose sign is determined by details of the magnetic order at the terminating surface. Manipulation of the surface termination and magnetic order can give rise to unidirectional 1D channels at hinges, surface steps and surface domain-walls\cite{yasuda17,khalaf-prb18,varnava-prb18}. Their protection from backscattering by the surface gap, the absence of external magnetic fields and the existence of robust and controllable quantum point junctions\cite{varnava-natcom21} make AFM TIs a prominent future material platform for quantum Hall physics. Furthermore, if all surfaces are gapped and have the same sign of the surface AHC (in a global sense), then the long-sought topological magnetoelectric effect can be observed, for which an applied electric field induces a parallel
magnetization and a magnetic field induces a parallel electric polarization with a quantized constant of proportionality given by $e^2/2h$.

The realization of magnetic TIs has become recently the focus of intense research\cite{tokura-nrp19,wang-scdir21,bernevig-nat22} with various candidates appearing in the literature\cite{otrokov-nat19,li-sc19,xu-prl19,hang-prx19,klimovskikh-npj20,ma-advmat20,xu-nat2020,rosa-npjqm2020,pierantozzie-pnas22}. The most prominent, MnBi$_2$Te$_4$\cite{otrokov-nat19,li-sc19}, is a layered tetradymite compound with an A-type antiferromagnetic (AFM) order, i.e., with magnetization uniform in-plane but alternating from plane to plane along the stacking direction. Control over the surface termination on thin films of MnBi$_2$Te$_4$ has resulted in the realization of high-temperature quantum anomalous Hall effect\cite{deng-sc20} as well as axion insulating states\cite{liu-natmat20}. However, with the nature and existence of surface gaps still disputed\cite{hang-prx19,wang21-scdir}, there is an evident need for improvement in the crystal quality as well as the search for new material candidates. 

\begin{figure*}[t]
\centering\includegraphics[width=7.0in]{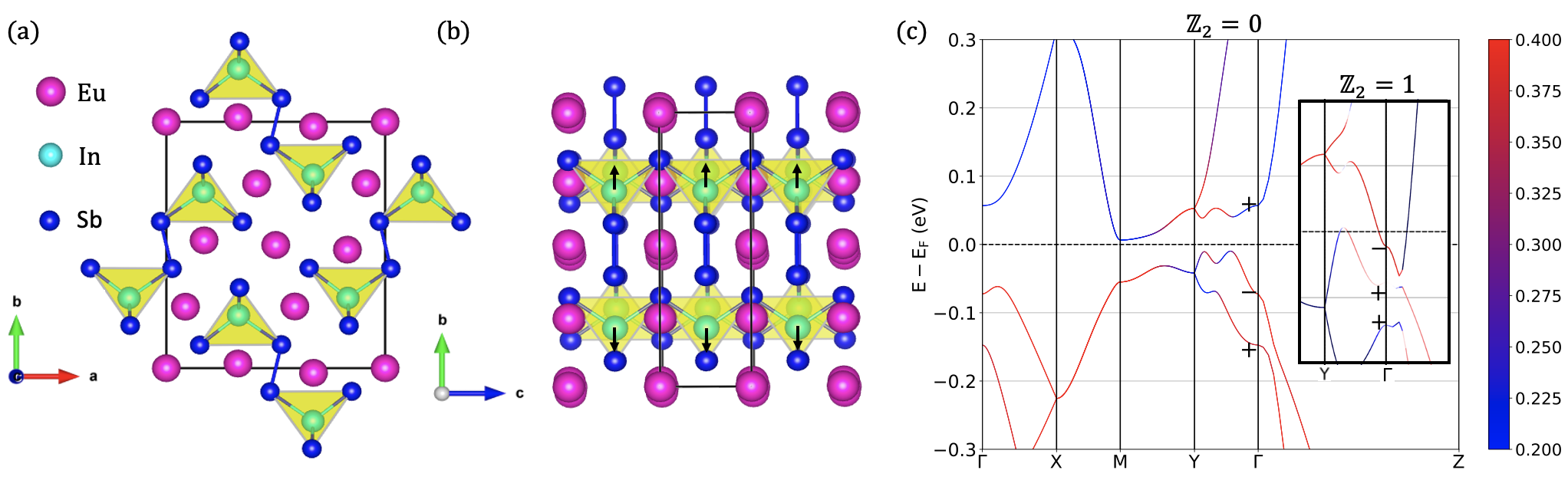}
\caption{Atomic and electronic structure of Eu$_5$In$_2$Sb$_6$. (a) The polyanions [In$_2$Sb$_6$]$^{-10}$ are comprised of pairs of approximate tetrahedra formed by Sb atoms and an In atom at their center. (b) Along the c-axis, the corner sharing tetrahedra form a quasi 1D chain. Eu and Sb$_1$ are located at integer layers while Sb$_2$, Sb$_3$ and In atoms are located at half-integer layers. Arrows indicate the direction along which the In atoms were displaced. (c) Electronic band structure of non-magnetic (Eu $f$-electrons in the core) Eu$_5$In$_2$Sb$_6$ in the presence of SOC. Inset shows the reproduced band structure from \citet{rosa-npjqm2020}, in which case the displacement of the In atoms causes the $\zt$ index to change.}
\label{fig:eu5in2sb6}
\end{figure*}

To this end, a class of materials known as Zintl compounds have attracted the attention of the community as promising in realizing magnetic TIs. These are defined as valence precise intermetallic phases in which electropositive cations donate electrons to covalently bonded polyanions~\cite{kauzlarich-tmd17}.  Numerous such compounds have already been synthesized and characterized, featuring a variety of interesting physical properties including superconductivity~\cite{deakin-jac02}, magnetoresistance~\cite{chan-cm97} and thermoelectricity~\cite{liu-jssc19}. Of particular interest in the search for magnetic TIs are the Zintls, in which the role of the cation is played by a divalent ${\rm Eu^{2+}}$ and the magnetism comes from the localized, spin-polarized $f$-orbital manifold. Examples of Eu-based Zintls appearing in the literature as candidate magnetic TIs include
$\mathrm{EuIn}_2\mathrm{As}_2$\cite{xu-prl19}, $\mathrm{EuSn}_2\mathrm{As}_2$\cite{hang-prx19},
$\mathrm{EuCd}_2\mathrm{As}_2$\cite{ma-advmat20}
Eu$_5$In$_2$Sb$_6$\cite{rosa-npjqm2020} and
$\mathrm{EuSn}_2\mathrm{P}_2$\cite{pierantozzie-pnas22}.

In this work, we gain insight into why Zintls are such a promising platform, and then use this  understanding to engineer magnetic TIs in Eu$_5$\textit{M$_2$X$_6$} (``526") Zintls where \textit{M}=Ga,In,Tl is a metal and \textit{X}=As,Sb,Bi is a pnictide. The insight relies on the notion of the complete electron transfer (CET) limit. In this limit, a Zintl should be a trivial insulator, but in fact there is electron competition between cations and polyanions. In this way, Zintl compounds form complex crystal and band structures that depend strongly on the geometry of the cations and polyanions. For topologically nontrivial Zintls, the presence of a spin-orbit coupling (SOC) induced bulk gap implies that the compound is away from the CET limit. The need to move away from the CET limit to obtain non-trivial topology, motivates us to use chemical substitution and structural perturbations to modify the crystal structure geometry and control the bands close to the Fermi level.

This insight was gained in part due to the discrepancy between our calculations on Eu$_5$In$_2$Sb$_6$ and those of \citet{rosa-npjqm2020}, who suggested that it has a non-trivial $\zt$ index. Our calculations on the compound clarify that it in fact it has a trivial $\zt$ index. After private communications with the authors of Ref.~\onlinecite{rosa-npjqm2020}, the discrepancy was identified to stem from a single transcription error in the cif file which was obtained from \citet{park-jmc02}. Having understood the geometrical implications of the discrepancy, we use Eu$_5$In$_2$Sb$_6$ as our starting point to computationally design new magnetic TIs in the 526 Eu-based family of Zintl compounds. We achieve this by means of chemical substitution and epitaxial as well as uniaxial strain to control both the bulk band inversions and bulk energy gaps. Specifically, our theory indicates that Eu$_5$Ga$_2$Sb$_6$, Eu$_5$Tl$_2$Sb$_6$ and Eu$_5$In$_2$Bi$_6$ form dynamically stable structures with non-trivial $\zt$ indices, and that epitaxial and uniaxial strain can be used to control their bulk energy gaps. However, the exact energy gaps are sensitive to the crystal structure and the magnetic configuration, which are not well established. 

With this motivation, we attempted the synthesis of the Eu$_5$Ga$_2$Sb$_6$ compound, but were not successful in isolating it as a pure phase. We then set out to explore the limit of stability of Ga substituted Eu$_5$(In$_{2-x}$Ga$_x$)$_2$Sb$_6$ and find an upper solubility limit of $x \approx 0.4$ based on polycrystalline synthesis. We find that the $c$ lattice constant contracts as the Ga concentration is increased, indicating a movement away from the CET limit in a way that is anticipated to push Eu$_5$In$_2$Sb$_6$ towards a magnetic topological insulating state.

%=================================================
\section{E\lowercase{u}$_5$I\lowercase{n}$_2$S\lowercase{b}$_6$} \seclabel{eu5in2sb6}

\subsection{Background}

The Zintl compound Eu$_5$In$_2$Sb$_6$ is a narrow-gap insulator that crystallizes in the orthorhombic space group (SG) Pbam (55)~\cite{park-jmc02}. The prominent structural features are [InSb$_4$] (approximate) tetrahedra that form pairs in the $ab$ plane through a short Sb-Sb bond, \fref{eu5in2sb6}(a). Following the Zintl concept, these tetrahedra pairs form covalently bonded polyanions [In$_2$Sb$_6$]$^{10-}$ while the divalent Eu atoms are dispersed between the polyanions providing the positive balancing charge 5[Eu$^{2+}$]. Along the $c$-axis the polyanions form quasi 1D chains via sharing tetrahedra corners (Sb$_1$), \fref{eu5in2sb6}(b). The compound has a layered structure along the $c$-axis with integer layers composed of Eu and Sb$_1$ and half-integer layers composed of In, Sb$_2$ and Sb$_3$, \fref{eu5in2sb6}(b). This picture implies that in the CET limit, close to the Fermi level, the valence bands will have Sb-p character while those in the conduction will have Eu-d character.

With respect to magnetism, the divalent Eu atoms have fully-polarized $4f$ orbitals with localized magnetic moments. The compound undergoes two magnetic transitions, at 14K and 7K. The moments are believed be antiferromagnetically aligned in the $ab$ plane, but the exact configuration still remains elusive. Importantly, the $4f$ states lie far from the Fermi level $E_{\rm F}$, in a narrow window $[-1.7{\rm eV},-1.3{\rm eV}]$. Thus treating them as core electrons has only a modest effect on the bulk bands near $E_{\rm F}$. 

The separation between the electronic and magnetic energy scales, the Zintl concept and the layered structure with all the Eu$^{2+}$ cations in the integer layers, point towards engineering of magnetic TIs. Namely, by setting the Eu $f$-electrons in the core we will consider the paramagnetic phase of Eu$^{2+}$, and use chemical substitution and structural perturbations such as epitaxial and uniaxial strain to tune the electronic band structure away from the CET limit and into the topological phase. In addition to tuning the topological phase, we will choose conditions under which the global energy gap is optimal. Since magnetism is not included in the calculation, the computational problem is greatly simplified. Recall that the topological index we are concerned with is the strong $\zt$ index of 3D TR-invariant insulators which can be determined using the Fu-Kane criterion\cite{fu-prb07} and the tools of topological quantum chemistry\cite{bradlyn-nat17}. As we explain in the Supplementary Information, the symmetry properties of the non-symmorphic SG 55 allow the determination of the topology just from the knowledge of inversion eigenvalues at $\Gamma$ and Z. In addition, the large energy gap at the $k_z=\pi$ plane, of the order of 1.5eV, implies that no band inversion will occur at the Z point. This offers an extra degree of robustness since we can focus on causing a band inversion at the $\Gamma$ point without worrying about band inversions at the Z point.

\subsection{The band structure}

In \fref{eu5in2sb6}(c) we show the band structure of Eu$_5$In$_2$Sb$_6$ in the case where the Eu $f$-electrons were set in the core. We use a color map to indicate the $p$-orbital character of the corresponding states, which serves as a visual cue for spotting band inversions. In the Supplementary Information, we compare \fref{eu5in2sb6}(c) with the band structure in the case where the putative A-type AFM configuration is assumed, to show that Eu magnetism only perturbs the states close to the Fermi level and therefore does not change the $\zt$ index. However, we note that the exact bulk band gap is sensitive to the magnetic configuration. In \fref{eu5in2sb6}(c), there is an evident band inversion, at the Y TRIM point, however the non-symmorphic symmetries force the same 4-dimensional representation, Y$_3$ +Y$_4$, at all states at Y so that Eu$_5$In$_2$Sb$_6$ is a trivial insulator. We also verify this using the Check Topological Mat. tool\cite{vergniory-nat19,vergniory-arxiv21} on the Bilbao crystallographic server\cite{aroyo-acs06,aroyo-zfk06,aroyo-bcc11}.

The inset of \fref{eu5in2sb6}(c) reproduces the band structure from \citet{rosa-npjqm2020}. In this case, due to a single transcription error\footnote{$y_{\rm In}=0.2419$ was used in their calculation while $y_{\rm In}=0.2149$ is reported  in \citet{park-jmc02}.}, the In atom was misplaced along the $b$-axis resulting in heavily distorted tetrahedra with two shorter (In-Sb$_1$) and two longer (In-Sb$_2$, In-Sb$_3$) bonds. An interpolation between the two structures in the absence of SOC, Supplementary Information, shows that as the tetrahedra become more heavily distorted, the character of the lowest conduction bands changes from Eu to In, signaling movement away from the CET limit. In this way, the overlap between conduction and valence bands increases. In the presence of SOC the increased overlap causes a band inversion at $\Gamma$ making the $\zt$ index non-trivial.  

Even if these results are negative with respect to Eu$_5$In$_2$Sb$_6$ being a magnetic TI, they nevertheless point to the highly tunable band structure of the compound. Indeed, the same kind of distortion of the tetrahedra would occur if we decreased the interlayer distance and/or increased the in-plane bond lengths. Physically, this could be achieved either by applying compressive uniaxial strain along the $c$-axis, or expanding the $a$ and $b$ lattice constants through epitaxial strain.

To understand why such geometric distortion can cause band inversions, note that in the CET limit, the charge migrates from the Eu atoms to the In$_2$Sb$_6$ polyanions. Since the compound has a layered structure with all Eu atoms on the integer layers, most of the accumulated charge will be in the half-integers layers and the system behaves like a dimerized chain along the $c$-axis. We should be careful with this analogy, however, since integer layers also include Sb$_1$ atom, so this only applies approximately. Compressing the dimerized chain should move us away-from the CET limit by increasing the overlap between conduction and valence bands. 

\begin{figure}
\includegraphics[width=3.5in]{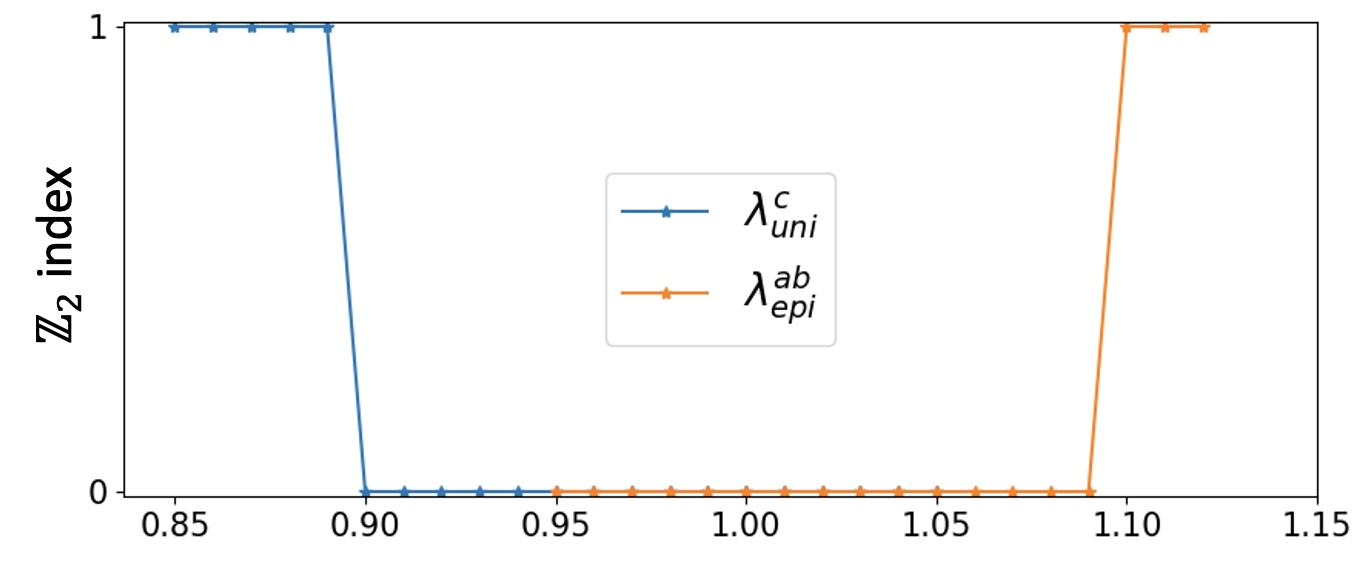}
\caption{The $\zt$ index of Eu$_5$In$_2$Sb$_6$ as a function of epitaxial and uniaxial strain. The $\zt$ index becomes non-trivial by either applying compressive uniaxial strain along $c$ or expansive epitaxial strain in the $ab$ plane. Strains of this magnitude are impractical but nevertheless provide insight into the mechanisms that drive Zintl compounds into topological phases.}
\label{fig:z2index}
\end{figure}

\begin{figure*}
\centering\includegraphics[width=7.0in]{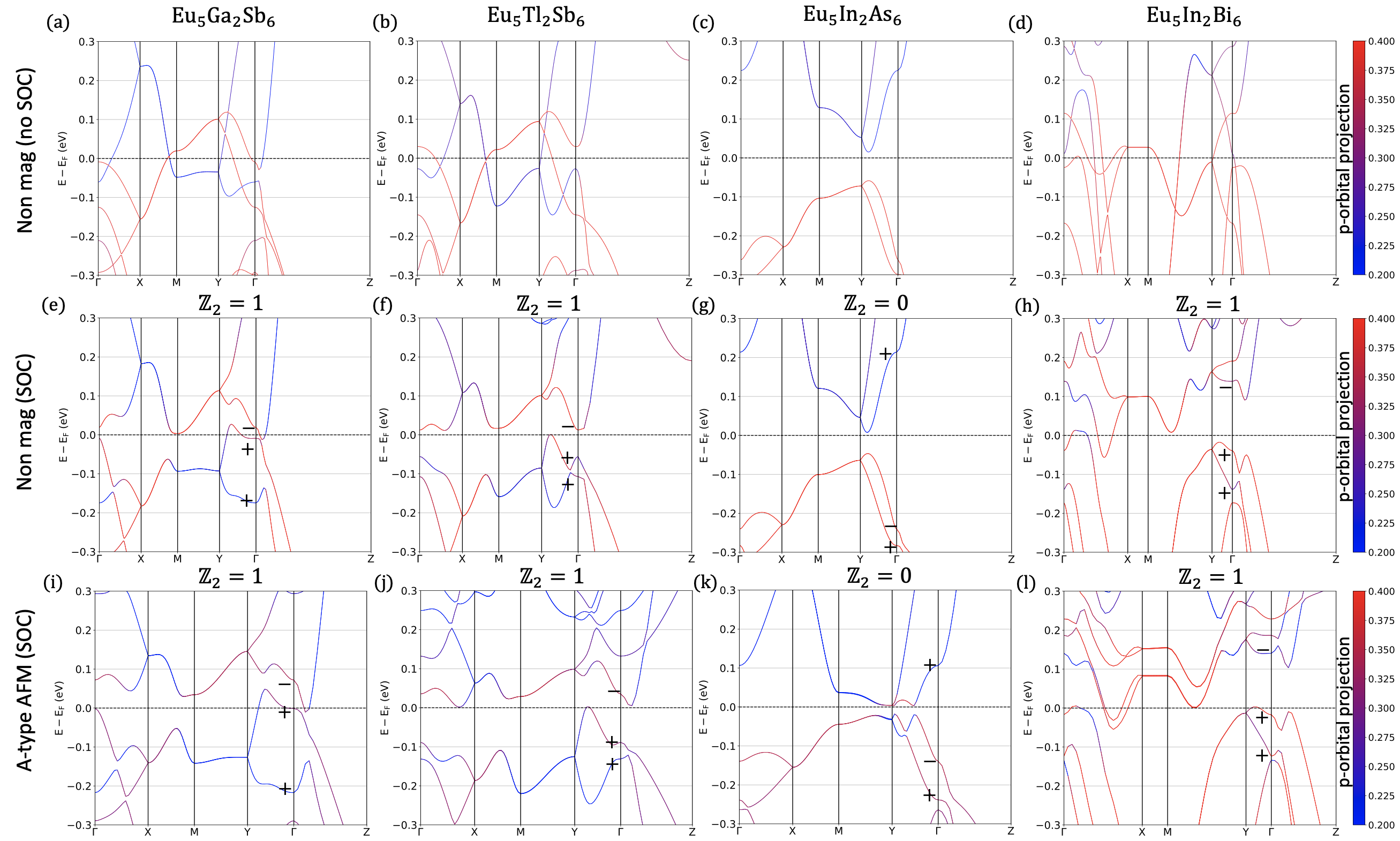}
\caption{Band structures (a)-(d) without magnetism and without SOC, (e)-(h) without magnetism and with SOC, (i)-(l) with an A-type AFM configuration and SOC for the four compounds obtained by isoelectronic substitution of Eu$_5$In$_2$Sb$_6$. Eu$_5$Ga$_2$Sb$_6$, Eu$_5$Tl$_2$Sb$_6$, Eu$_5$In$_2$Bi$_6$ are topologically non-trivial while  Eu$_5$In$_2$As$_6$ is trivial.}
\label{fig:bands}
\end{figure*}

Using DFT, we can simulate the effect of uniaxial strain along the $c$-axis by multiplying the lattice constant $c_0$ by a coefficient $\lambda^c_\text{uni}$ so that the strained lattice constant $c$ is given by $c=\lambda^c_\text{uni} c_0$. We then relax the structure but keep $c$ fixed, see Methods. Similarly, for epitaxial strain, we keep the in-plane lattice constants $a=\lambda^{ab}_\text{epi} a_0$, $b=\lambda^{ab}_\text{epi} b_0$ fixed and relax the structure. \fref{z2index} shows that a $10\%$ compressive uniaxial strain or a $10\%$ expansive epitaxial converts the $\zt$ index from trivial to non-trivial. Of course, such extreme strains are impractical for any real applications, but these calculations provide insight into mechanisms that can drive Zintl compounds into topological phases. In the Supplementary Information, we plot the evolution of the band structure as a function of the strain coefficients, which shows that the effect of displacing the In atom along the $b$-axis is very similar to applying epitaxial and uniaxial strain.

\section{Tuning band structure properties} \seclabel{prediction}

\subsection{Via chemical substitution}

The sensitivity of the band structure to the position of the In atom motivates us to consider its isoelectronic substitution with Ga or Tl. In this way, we can apply chemical pressure while preserving the other properties. After relaxing the substituted structures (both the internal coordinates and lattice constants), we find that the system remains in SG 55 and the phonon frequencies all remain positive, indicating that the structures are dynamically stable, Supplementary Information. In the absence of SOC, \fref{bands}(a),(b), the overlap between valence and conduction bands is increased compared to Eu$_5$In$_2$Sb$_6$, indicating that Eu$_5$Ga$_2$Sb$_6$ and Eu$_5$Tl$_2$Sb$_6$ are further away from the CET limit. Just like the case of the displaced In atoms, the lowest conduction states have mostly In-Sb$_1$ character and the highest valence states have mostly  Sb$_2$-Sb$_3$ character. We note, however, that in the Ga and Tl substituted compounds the tetrahedra environment is not distorted. When SOC is included a band inversion at $\Gamma$ occurs, exchanging positive and negative parity eigenvalues, \fref{bands}(e),(f), resulting in a non-trivial $\zt$ index. Here considering the putative A-type AFM configuration, \fref{bands}(i),(j), does not change the topology and has a small effect on the bulk bands, moving the compounds further away from the CET limit.

Another way to alter the tetrahedra while preserving the chemical properties is to substitute the Sb in Eu$_5$In$_2$Sb$_6$ with either As or Bi. Since Sb atoms are three times more abundant than the In atoms and play a central role in determining the polyanion structure, such a substitution has much more severe effect than substitution by Ga or Tl. Fortunately, both  Eu$_5$In$_2$As$_6$\cite{childs-jssc19} and Eu$_5$In$_2$Bi$_6$\cite{radzieowski-mcf20} have been recently synthesized, so their crystallographic structures are known.  We find that Eu$_5$In$_2$As$_6$ is closer to the CET limit, \fref{bands}(c), and therefore is trivial, \fref{bands}(g). Instead Eu$_5$In$_2$Bi$_6$ is away from the CET limit, \fref{bands}(d), and has a non-trivial $\zt$ index, \fref{bands}(h). We also verify that the A-type AFM configuration does have a significant effect on the bulk bands and no effect on the topology \fref{bands}(k),(l).

\begin{table}[ht]
\centering
\caption{Key properties of the 526 Eu-based Zintls. } \label{tab:table1}
\begin{tabular}{p{1.7cm}p{0.8cm}p{0.8cm}p{0.8cm}p{0.4cm}p{1.0cm}p{1.0cm}p{1.0cm}}
\hline\hline 
Compound & a(\AA) & b(\AA) & c(\AA) & $\zt$ & Exists? & DG & IG \\
         &  &  &  &  & & (meV) & (meV) \\
\hline
Eu$_5$In$_2$Sb$_6$ & 12.51 & 14.58 & 4.62  & 0 & \checkmark & 45 & 23 \\
Eu$_5$Ga$_2$Sb$_6$ & 12.47 & 14.32 & 4.54  & 1 &  \text{\sffamily X} & 16 & -39 \\
Eu$_5$Tl$_2$Sb$_6$ & 12.48 & 14.67 & 4.70  & 1 &  \text{\sffamily X}& 70 & 8  \\
Eu$_5$In$_2$As$_6$ & 11.87 & 13.78 & 4.35  & 0 &  \checkmark & 54  & 54 \\
Eu$_5$In$_2$Bi$_6$ &  7.77 & 24.08 & 4.70  & 1 &  \checkmark & 77  & -69 \\
\end{tabular}
\end{table}

\begin{figure*}
\centering\includegraphics[width=7.0in]{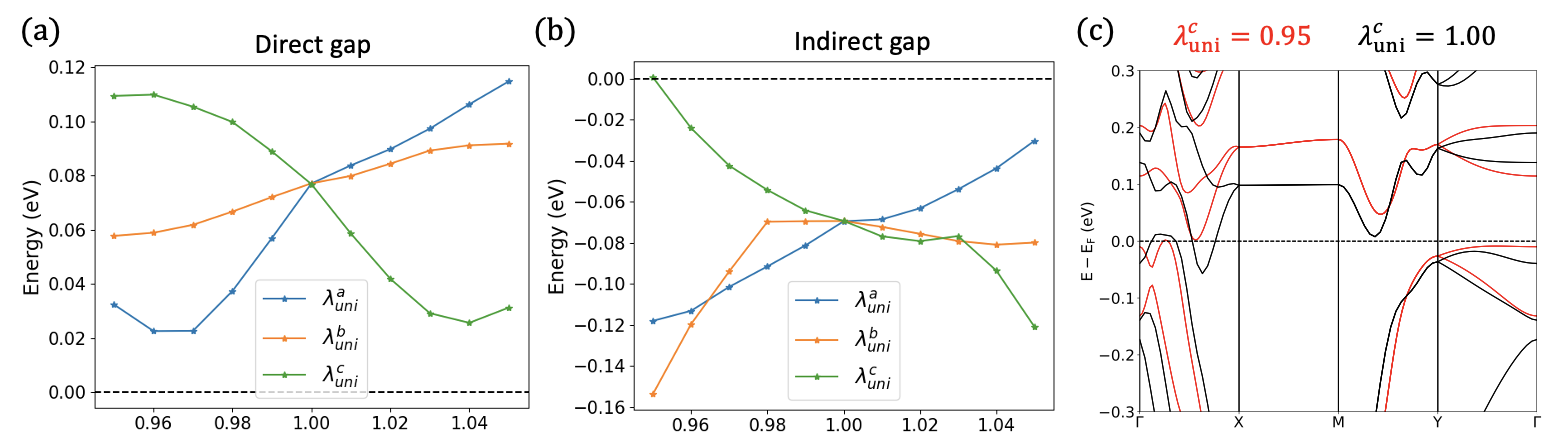}
\caption{Effect of uniaxial strain on (a) direct and  (b) indirect gap. (e) Comparison between band structures with $\lambda^c_\text{uni}=1.00$ and $\lambda^c_\text{uni}=0.95$.}
\label{fig:eu5in2bi6}
\end{figure*}

Table \ref{tab:table1} contains a summary of some of the key features of the 526 compounds discussed. All compounds are in the same space group as the Eu$_5$In$_2$Sb$_6$ parent compound. However, Eu$_5$In$_2$Bi$_6$ follows a different structure type with modified $a$ and $b$ lattice constants, and the Bi-Bi bond connecting two tetrahedra is now along $a$-axis instead of the $b$-axis as in the other compounds. With respect to the direct energy gap, the compounds with heavier elements such as Tl or Bi have the largest gap, in part due to the enhanced effect of SOC. In contrast, the indirect or global gap of all nontrivial candidates is negative, with the exception of Eu$_5$Tl$_2$Sb$_6$ which has a small indirect gap. Unfortunately, the toxic nature of Tl tends to preclude practical applications. It is therefore desirable to modify the band structures so as to increase the indirect band gap and remove the electron and hole pockets, \fref{bands}.

In \sref{exp} we will discuss the experimental progress towards realizing the magnetic topological insulator in Eu$_5$Ga$_2$Sb$_6$. 

\subsection{Via strain engineering} \seclabel{engineering}

We have seen that we can use uniaxial and epitaxial strain to change the topological index by moving away from the CET limit. An important factor in this kind of engineering was the insensitivity of these systems to band inversions other than at the $\Gamma$ and Z points due to the non-symmorphic nature of the space group. In this section, we show that moving away from the CET limit can also increase the direct and indirect band gaps. This is much less obvious than changing the topological index, as the gaps are determined by the global band structure properties, i.e., in the whole Brillouin zone. We use Eu$_5$In$_2$Bi$_6$ as a case study since the presence of Bi gives rise to larger SOC-induced gaps.

\fref{eu5in2bi6}(a),(b), show that uniaxial strain with $\lambda^a_\text{uni}<1.00$ or $\lambda^c_\text{uni}>1.00$ enhances the direct and indirect gaps. These kinds of uniaxial strain move the compound away from the CET limit by increasing the band overlap (in the absence of SOC). In \fref{eu5in2bi6}(c) we compare the band structures with $\lambda^c_\text{uni}=1.00$ and $\lambda^c_\text{uni}=0.95$ to show that uniaxial strain can be used to remove electron and hole pockets. The same trends occur with epitaxial strain as we discuss in the Supplementary Information. Finally, we should note that the direct and indirect energy gaps will be sensitive to the exact magnetic configuration, but the trends should not depend on it.

\section{Experimental progress}\seclabel{exp}

\begin{figure}
\includegraphics[width=3.5in]{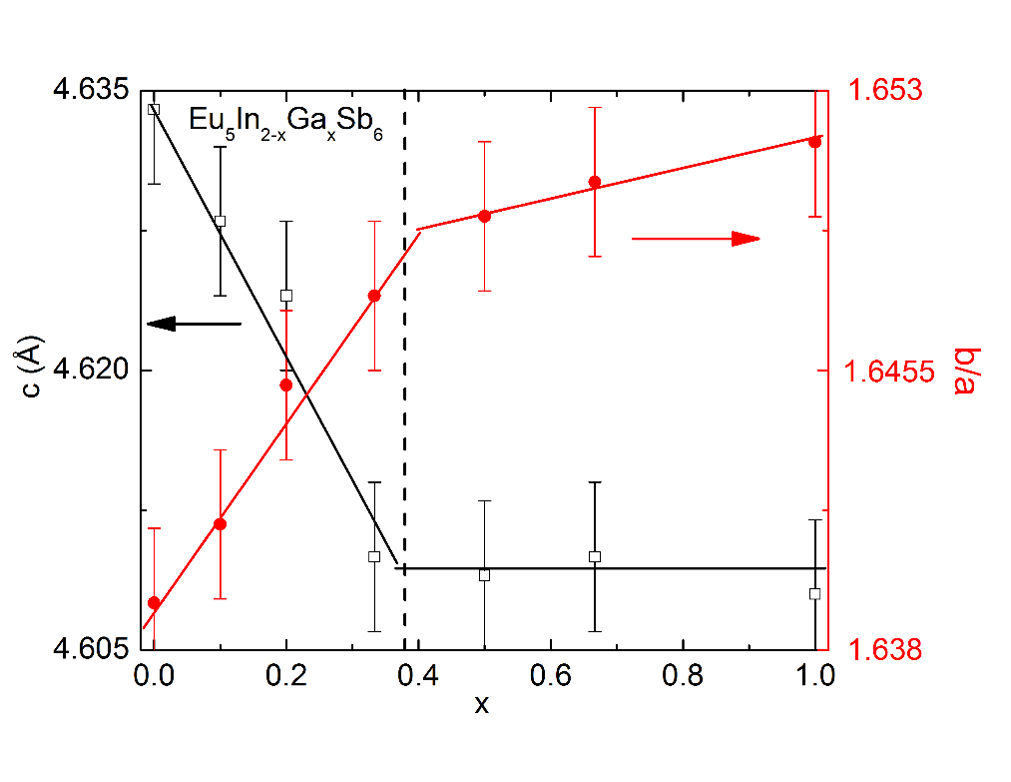}
\caption{The $c$ axis and $b/a$ ratio of polycrystalline Eu$_5$In$_{2-x}$Ga$_x$Sb$_6$ show a systematic evolution with Ga substitution up to $x \approx 0.4$, the apparent solid solubility limit. Contraction of the $c$ lattice constant with increased Ga concentration indicates movement away from the CET limit, in agreement with the DFT prediction.}
\label{fig:exp_in_ga}
\end{figure}

Motivated by the strong theoretical predictions of magnetic topological insulating states in Eu$_5$Ga$_2$Sb$_6$, we attempted its preparation following known synthetic recipes for Eu$_5$In$_2$Sb$_6$\cite{park-jmc02,subbarao-jac16,rosa-npjqm2020}. In agreement with prior reports\cite{subbarao-jac16}, we were unable to find synthetic conditions in which to stabilize the Ga phase. Prior work on topological insulators has shown that solid solutions are a viable avenue to tune across topological phase transitions, e.g. Bi$_{1-x}$Sb$_x$\cite{teo-prb08,hsieh-nat08} and TlBiS$_{2-x}$Se$_x$\cite{sato-nat11}. With this in mind, we utilized polycrystalline synthesis to find the solid solubility limit of Ga in Eu$_5$In$_{2-x}$Ga$_x$Sb$_6$. \fref{exp_in_ga} summarizes our findings. As the structure of Eu$_5$In$_2$Sb$_6$ is orthorhombic, with chains of (In/Ga)Sb$_4$ tetrahedra aligned along the $c$ axis, we find an expected contraction along the $c$ axis, and a smaller change in the $ab$ plane where the tetrahedral slightly contract, but retain optimum bonding environments for Eu. Specifically the $c$ lattice parameter decreases from $4.634\textrm{\AA}$ to a minimum of $4.608\textrm{\AA}$. The $a$ lattice parameter decreases slightly from $12.530\textrm{\AA}$ to $12.520\textrm{\AA}$ over the same range, while $b$ is unchanged within error ($14.584\textrm{\AA}$ at $x = 0$). This contraction of the $c$ axis should push the material further away from the CET limit and towards a topological state. Further evidence for successful substitution comes from the monotonic increase in the $b/a$ ratio over the same composition range, implying a change in the relative shape in the $ab$ plane. This demonstrates that partial replacement of In with Ga is possible, and motivates further work to both extend the phase stability line, and determine whether there is experimentally a change in topology in the chemically accessible range. 

In addition, future work includes the synthesis of Eu$_5$In$_2$Bi$_6$ single crystals utilizing the flux technique. Prior works suggests the existence of Eu$_5$In$_2$Bi$_6$ in polycrystalline form, but as a minority phase, and phase purity could not be achieved\cite{radzieowski-mcf20}. By utilizing the flux technique we hope to segregate the thermodynamically unstable, Eu$_5$In$_2$Bi$_6$ phase in the single-crystal form. The single crystals would also allow us to understand the nature of transport and magnetism in Eu$_5$In$_2$Bi$_6$ and allow us to explore new avenues to the discovery of magnetic topological states of matter.

\section{Summary and conclusions}

The Zintl concept is an example of a chemical concept that provides a bridge between the structural and electronic properties for a particular class of compounds. This is especially interesting for the engineering of topological insulators, which require inverted insulating gaps. In addition, the large number of Zintl compounds that have been synthesized means that there is a big enough search space to allow optimization of desired properties such as bulk and surface gaps. DFT is an indispensable tool in this search, since it allows accurate prediction of material properties. 

In this work, we gain deeper understanding into how structural and electronic properties interact in the 526 family of Eu-based Zintl compounds. We use the Zintl concept to relate the layered structure to an ionic chain along $c$. By realizing that we need to move away from the CET limit, we find ways to change the topology and remove electron and hole pockets. These insights will be applicable in various other systems of Zintl compounds. For example, it was recently predicted that the Eu-based Zintl $\mathrm{EuCd}_2\mathrm{Sb}_2$\cite{runhan21} becomes a 2D AFM TI under tensile strain. This is explained by the same mechanism we described here as all divalent Eu atoms are located in a single layer as for the 526 Zintls.

Finally we note that, due to the highly tunable energy gap in Zintls, whether the proposed candidates will have gapped bulk and surface gaps, depends will depend sensitively on the crystal structure, the magnetic configuration. Therefore, the aim of this work is to show and explain trends and provide motivation for further investigation and not to make quantitative prediction.

\section{Methods} \seclabel{methods}
Density functional theory\cite{hohenberg-pr64,kohn-pr65} (DFT) based first principles calculations were performed using  the projector augmented-wave (PAW) method as implemented in the
VASP code\cite{kresse-prb96,kresse-prb99}. We used the PBE exchange-correlation functional as parametrized by Perdew-Burke-Ernzerhof\cite{perdew-prl96}. The Brillouin zone sampling was performed by using a $11 \times 3 \times 15$ k-mesh for Eu$_5$In$_2$Bi$_6$ and $5 \times 5 \times 15$ for all other compounds. The energy cutoff is chosen 1.5 times as large as the values recommended for the relevant PAW pseudopotentials. Spin-orbit coupling (SOC) was included self-consistently. The Eu $4f$ states were treated by employing the GGA+U approach with the U value set to 5.0eV. 

Structure relaxation calculations were performed using ISIF=3 tag for which forces and the stress tensor are calculated and all degrees of freedom are relaxed. To model the effect of uniaxial strain, one of the lattice constants was fixed and all other degrees of freedom were allowed to relax. Similarly we modeled epitaxial strain by fixing two of the lattice constants and allowing all other degrees of freedom to relax. Phonon calculations were carried out using the PHONOPY package\cite{phonopy}. Irreducible representations and their traces were calculated using Irvsp\cite{gao-cpc21}. Band structure plots were obtained using PyProcar\cite{pyprocar}.

Polycrystalline Eu$_5$In$_{2-x}$Ga$_x$Sb$_6$ were prepared in a manner similar to that previously reported for Eu$_5$In$_2$Sb$_6$, with equimolal replacement of In with Ga\cite{park-jmc02,subbarao-jac16}. Powder x-ray diffraction patterns were collected over an angular range of $5-90^{\circ} 2\theta$ on a Bruker D8 focus equipped with a LynxEye detector, 1mm incident slit, soller slits, and an air antiscatter shield, and indexed using Lebail refinements as implemented in Bruker TOPAS.

\section{Acknowledgements} \seclabel{acknowledgements}
We acknowledge constructive discussions with N. Peter Armitage, Collin Broholm, Vincent Morano, Priscila Rosa, Yuanfeng Xu,  Zhijun Wang, Andrei Bernevig. N.V. thanks Sobhit Singh for teaching him how to perform phonon calculations and providing the VASP code to perform epitaxial and uniaxial strain calculations. This work was supported as part of the Institute for Quantum Matter and Energy Frontier Research Center, funded by the U.S. Department of Energy, Office of Science, Office of Basic Energy Sciences, under Award DE-SC0019331. 

\section{Competing interests} 
The authors declare no competing interests.

\section{Author Contributions} 
N.V. performed the DFT calculations and analysis under the supervision of D.V.. Experimental work was done by T.B. under the supervision of T.M.. N.V. wrote the manuscript with input from all authors. 

\bibliography{pap}

\begin{thebibliography}{52}%
\makeatletter
\providecommand \@ifxundefined [1]{%
 \@ifx{#1\undefined}
}%
\providecommand \@ifnum [1]{%
 \ifnum #1\expandafter \@firstoftwo
 \else \expandafter \@secondoftwo
 \fi
}%
\providecommand \@ifx [1]{%
 \ifx #1\expandafter \@firstoftwo
 \else \expandafter \@secondoftwo
 \fi
}%
\providecommand \natexlab [1]{#1}%
\providecommand \enquote  [1]{``#1''}%
\providecommand \bibnamefont  [1]{#1}%
\providecommand \bibfnamefont [1]{#1}%
\providecommand \citenamefont [1]{#1}%
\providecommand \href@noop [0]{\@secondoftwo}%
\providecommand \href [0]{\begingroup \@sanitize@url \@href}%
\providecommand \@href[1]{\@@startlink{#1}\@@href}%
\providecommand \@@href[1]{\endgroup#1\@@endlink}%
\providecommand \@sanitize@url [0]{\catcode `\\12\catcode `\$12\catcode
  `\&12\catcode `\#12\catcode `\^12\catcode `\_12\catcode `\%12\relax}%
\providecommand \@@startlink[1]{}%
\providecommand \@@endlink[0]{}%
\providecommand \url  [0]{\begingroup\@sanitize@url \@url }%
\providecommand \@url [1]{\endgroup\@href {#1}{\urlprefix }}%
\providecommand \urlprefix  [0]{URL }%
\providecommand \Eprint [0]{\href }%
\providecommand \doibase [0]{http://dx.doi.org/}%
\providecommand \selectlanguage [0]{\@gobble}%
\providecommand \bibinfo  [0]{\@secondoftwo}%
\providecommand \bibfield  [0]{\@secondoftwo}%
\providecommand \translation [1]{[#1]}%
\providecommand \BibitemOpen [0]{}%
\providecommand \bibitemStop [0]{}%
\providecommand \bibitemNoStop [0]{.\EOS\space}%
\providecommand \EOS [0]{\spacefactor3000\relax}%
\providecommand \BibitemShut  [1]{\csname bibitem#1\endcsname}%
\let\auto@bib@innerbib\@empty
%</preamble>
\bibitem [{\citenamefont {Hasan}\ and\ \citenamefont
  {Kane}(2010)}]{hasan-rmp10}%
  \BibitemOpen
  \bibfield  {author} {\bibinfo {author} {\bibfnamefont {M.~Z.}\ \bibnamefont
  {Hasan}}\ and\ \bibinfo {author} {\bibfnamefont {C.~L.}\ \bibnamefont
  {Kane}},\ }\bibfield  {title} {\enquote {\bibinfo {title} {Colloquium:
  Topological insulators},}\ }\href {\doibase 10.1103/RevModPhys.82.3045}
  {\bibfield  {journal} {\bibinfo  {journal} {Rev. Mod. Phys.}\ ,\ \bibinfo
  {pages} {3045}} (\bibinfo {year} {2010})}\BibitemShut {NoStop}%
\bibitem [{\citenamefont {Qi}\ and\ \citenamefont {Zhang}(2011)}]{qi-prm11}%
  \BibitemOpen
  \bibfield  {author} {\bibinfo {author} {\bibfnamefont {Xiao-Liang}\
  \bibnamefont {Qi}}\ and\ \bibinfo {author} {\bibfnamefont {Shou-Cheng}\
  \bibnamefont {Zhang}},\ }\bibfield  {title} {\enquote {\bibinfo {title}
  {Topological insulators and superconductors},}\ }\href {\doibase
  10.1103/RevModPhys.83.1057} {\bibfield  {journal} {\bibinfo  {journal} {Rev.
  Mod. Phys.}\ }\textbf {\bibinfo {volume} {83}},\ \bibinfo {pages}
  {1057--1110} (\bibinfo {year} {2011})}\BibitemShut {NoStop}%
\bibitem [{\citenamefont {Qi}\ \emph {et~al.}(2008)\citenamefont {Qi},
  \citenamefont {Hughes},\ and\ \citenamefont {Zhang}}]{qi-prb08}%
  \BibitemOpen
  \bibfield  {author} {\bibinfo {author} {\bibfnamefont {X.~L.}\ \bibnamefont
  {Qi}}, \bibinfo {author} {\bibfnamefont {T.~L.}\ \bibnamefont {Hughes}}, \
  and\ \bibinfo {author} {\bibfnamefont {S.~C.}\ \bibnamefont {Zhang}},\
  }\bibfield  {title} {\enquote {\bibinfo {title} {Topological field theory of
  time-reversal invariant insulators},}\ }\href {\doibase
  10.1103/PhysRevB.78.195424} {\bibfield  {journal} {\bibinfo  {journal} {Phys.
  Rev. B}\ }\textbf {\bibinfo {volume} {78}},\ \bibinfo {pages} {195424}
  (\bibinfo {year} {2008})}\BibitemShut {NoStop}%
\bibitem [{\citenamefont {Essin}\ \emph {et~al.}(2009)\citenamefont {Essin},
  \citenamefont {Moore},\ and\ \citenamefont {Vanderbilt}}]{essin-prl09+e}%
  \BibitemOpen
  \bibfield  {author} {\bibinfo {author} {\bibfnamefont {Andrew~M.}\
  \bibnamefont {Essin}}, \bibinfo {author} {\bibfnamefont {Joel~E.}\
  \bibnamefont {Moore}}, \ and\ \bibinfo {author} {\bibfnamefont {David}\
  \bibnamefont {Vanderbilt}},\ }\bibfield  {title} {\enquote {\bibinfo {title}
  {Magnetoelectric polarizability and axion electrodynamics in crystalline
  insulators},}\ }\href {\doibase 10.1103/PhysRevLett.102.146805} {\bibfield
  {journal} {\bibinfo  {journal} {Phys. Rev. Lett.}\ }\textbf {\bibinfo
  {volume} {102}},\ \bibinfo {eid} {146805} (\bibinfo {year} {2009})},\
  \bibinfo {note} {see also Erratum: Phys. Rev. Lett. {\bf 103}, 259902(E)
  (2009).}\BibitemShut {Stop}%
\bibitem [{\citenamefont {Varnava}\ \emph {et~al.}(2020)\citenamefont
  {Varnava}, \citenamefont {Souza},\ and\ \citenamefont
  {Vanderbilt}}]{varnava-prb20}%
  \BibitemOpen
  \bibfield  {author} {\bibinfo {author} {\bibfnamefont {Nicodemos}\
  \bibnamefont {Varnava}}, \bibinfo {author} {\bibfnamefont {Ivo}\ \bibnamefont
  {Souza}}, \ and\ \bibinfo {author} {\bibfnamefont {David}\ \bibnamefont
  {Vanderbilt}},\ }\bibfield  {title} {\enquote {\bibinfo {title} {Axion
  coupling in the hybrid wannier representation},}\ }\href {\doibase
  10.1103/PhysRevB.101.155130} {\bibfield  {journal} {\bibinfo  {journal}
  {Phys. Rev. B}\ }\textbf {\bibinfo {volume} {101}},\ \bibinfo {pages}
  {155130} (\bibinfo {year} {2020})}\BibitemShut {NoStop}%
\bibitem [{\citenamefont {Yasuda}\ \emph {et~al.}(2017)\citenamefont {Yasuda},
  \citenamefont {Mogi}, \citenamefont {Yoshimi}, \citenamefont {Tsukazaki},
  \citenamefont {Takahashi}, \citenamefont {Kawasaki}, \citenamefont {Kagawa},\
  and\ \citenamefont {Tokura}}]{yasuda17}%
  \BibitemOpen
  \bibfield  {author} {\bibinfo {author} {\bibfnamefont {K.}~\bibnamefont
  {Yasuda}}, \bibinfo {author} {\bibfnamefont {M.}~\bibnamefont {Mogi}},
  \bibinfo {author} {\bibfnamefont {R.}~\bibnamefont {Yoshimi}}, \bibinfo
  {author} {\bibfnamefont {A.}~\bibnamefont {Tsukazaki}}, \bibinfo {author}
  {\bibfnamefont {K.~S.}\ \bibnamefont {Takahashi}}, \bibinfo {author}
  {\bibfnamefont {M.}~\bibnamefont {Kawasaki}}, \bibinfo {author}
  {\bibfnamefont {F.}~\bibnamefont {Kagawa}}, \ and\ \bibinfo {author}
  {\bibfnamefont {Y.}~\bibnamefont {Tokura}},\ }\bibfield  {title} {\enquote
  {\bibinfo {title} {Quantized chiral edge conduction on domain walls of a
  magnetic topological insulator},}\ }\href {\doibase 10.1126/science.aan5991}
  {\bibfield  {journal} {\bibinfo  {journal} {Science}\ }\textbf {\bibinfo
  {volume} {358}},\ \bibinfo {pages} {1311--1314} (\bibinfo {year} {2017})},\
  \Eprint
  {http://arxiv.org/abs/https://www.science.org/doi/pdf/10.1126/science.aan5991}
  {https://www.science.org/doi/pdf/10.1126/science.aan5991} \BibitemShut
  {NoStop}%
\bibitem [{\citenamefont {Khalaf}(2018)}]{khalaf-prb18}%
  \BibitemOpen
  \bibfield  {author} {\bibinfo {author} {\bibfnamefont {E.}~\bibnamefont
  {Khalaf}},\ }\bibfield  {title} {\enquote {\bibinfo {title} {Higher-order
  topological insulators and superconductors protected by inversion
  symmetry},}\ }\href {\doibase 10.1103/PhysRevB.97.205136} {\bibfield
  {journal} {\bibinfo  {journal} {Phys. Rev. B}\ }\textbf {\bibinfo {volume}
  {97}},\ \bibinfo {pages} {205136} (\bibinfo {year} {2018})}\BibitemShut
  {NoStop}%
\bibitem [{\citenamefont {Varnava}\ and\ \citenamefont
  {Vanderbilt}(2018)}]{varnava-prb18}%
  \BibitemOpen
  \bibfield  {author} {\bibinfo {author} {\bibfnamefont {Nicodemos}\
  \bibnamefont {Varnava}}\ and\ \bibinfo {author} {\bibfnamefont {David}\
  \bibnamefont {Vanderbilt}},\ }\bibfield  {title} {\enquote {\bibinfo {title}
  {Surfaces of axion insulators},}\ }\href {\doibase
  10.1103/PhysRevB.98.245117} {\bibfield  {journal} {\bibinfo  {journal} {Phys.
  Rev. B}\ }\textbf {\bibinfo {volume} {98}},\ \bibinfo {pages} {245117}
  (\bibinfo {year} {2018})}\BibitemShut {NoStop}%
\bibitem [{\citenamefont {Varnava}\ \emph {et~al.}(2021)\citenamefont
  {Varnava}, \citenamefont {Wilson}, \citenamefont {Pixley},\ and\
  \citenamefont {Vanderbilt}}]{varnava-natcom21}%
  \BibitemOpen
  \bibfield  {author} {\bibinfo {author} {\bibfnamefont {Nicodemos}\
  \bibnamefont {Varnava}}, \bibinfo {author} {\bibfnamefont {Justin~H.}\
  \bibnamefont {Wilson}}, \bibinfo {author} {\bibfnamefont {J.~H.}\
  \bibnamefont {Pixley}}, \ and\ \bibinfo {author} {\bibfnamefont {David}\
  \bibnamefont {Vanderbilt}},\ }\bibfield  {title} {\enquote {\bibinfo {title}
  {Controllable quantum point junction on the surface of an antiferromagnetic
  topological insulator},}\ }\href {\doibase 10.1038/s41467-021-24276-5}
  {\bibfield  {journal} {\bibinfo  {journal} {Nature Communications}\ }\textbf
  {\bibinfo {volume} {12}},\ \bibinfo {pages} {3998} (\bibinfo {year}
  {2021})}\BibitemShut {NoStop}%
\bibitem [{\citenamefont {Tokura}\ \emph {et~al.}(2019)\citenamefont {Tokura},
  \citenamefont {Yasuda},\ and\ \citenamefont {Tsukazaki}}]{tokura-nrp19}%
  \BibitemOpen
  \bibfield  {author} {\bibinfo {author} {\bibfnamefont {Yoshinori}\
  \bibnamefont {Tokura}}, \bibinfo {author} {\bibfnamefont {Kenji}\
  \bibnamefont {Yasuda}}, \ and\ \bibinfo {author} {\bibfnamefont {Atsushi}\
  \bibnamefont {Tsukazaki}},\ }\bibfield  {title} {\enquote {\bibinfo {title}
  {Magnetic topological insulators},}\ }\href {\doibase
  10.1038/s42254-018-0011-5} {\bibfield  {journal} {\bibinfo  {journal} {Nature
  Reviews Physics}\ }\textbf {\bibinfo {volume} {1}},\ \bibinfo {pages}
  {126--143} (\bibinfo {year} {2019})}\BibitemShut {NoStop}%
\bibitem [{\citenamefont {Wang}\ \emph
  {et~al.}(2021{\natexlab{a}})\citenamefont {Wang}, \citenamefont {Ge},
  \citenamefont {Li}, \citenamefont {Liu}, \citenamefont {Xu},\ and\
  \citenamefont {Wang}}]{wang-scdir21}%
  \BibitemOpen
  \bibfield  {author} {\bibinfo {author} {\bibfnamefont {Pinyuan}\ \bibnamefont
  {Wang}}, \bibinfo {author} {\bibfnamefont {Jun}\ \bibnamefont {Ge}}, \bibinfo
  {author} {\bibfnamefont {Jiaheng}\ \bibnamefont {Li}}, \bibinfo {author}
  {\bibfnamefont {Yanzhao}\ \bibnamefont {Liu}}, \bibinfo {author}
  {\bibfnamefont {Yong}\ \bibnamefont {Xu}}, \ and\ \bibinfo {author}
  {\bibfnamefont {Jian}\ \bibnamefont {Wang}},\ }\bibfield  {title} {\enquote
  {\bibinfo {title} {Intrinsic magnetic topological insulators},}\ }\href
  {\doibase https://doi.org/10.1016/j.xinn.2021.100098} {\bibfield  {journal}
  {\bibinfo  {journal} {The Innovation}\ }\textbf {\bibinfo {volume} {2}},\
  \bibinfo {pages} {100098} (\bibinfo {year} {2021}{\natexlab{a}})}\BibitemShut
  {NoStop}%
\bibitem [{\citenamefont {Bernevig}\ \emph {et~al.}(2022)\citenamefont
  {Bernevig}, \citenamefont {Felser},\ and\ \citenamefont
  {Beidenkopf}}]{bernevig-nat22}%
  \BibitemOpen
  \bibfield  {author} {\bibinfo {author} {\bibfnamefont {B.~Andrei}\
  \bibnamefont {Bernevig}}, \bibinfo {author} {\bibfnamefont {Claudia}\
  \bibnamefont {Felser}}, \ and\ \bibinfo {author} {\bibfnamefont {Haim}\
  \bibnamefont {Beidenkopf}},\ }\bibfield  {title} {\enquote {\bibinfo {title}
  {Progress and prospects in magnetic topological materials},}\ }\href
  {\doibase 10.1038/s41586-021-04105-x} {\bibfield  {journal} {\bibinfo
  {journal} {Nature}\ }\textbf {\bibinfo {volume} {603}},\ \bibinfo {pages}
  {41--51} (\bibinfo {year} {2022})}\BibitemShut {NoStop}%
\bibitem [{\citenamefont {Otrokov}\ \emph {et~al.}(2019)\citenamefont
  {Otrokov}, \citenamefont {Klimovskikh}, \citenamefont {Bentmann},
  \citenamefont {Estyunin}, \citenamefont {Zeugner}, \citenamefont {Aliev},
  \citenamefont {Ga{\ss}}, \citenamefont {Wolter}, \citenamefont {Koroleva},
  \citenamefont {Shikin}, \citenamefont {Blanco-Rey}, \citenamefont {Hoffmann},
  \citenamefont {Rusinov}, \citenamefont {Vyazovskaya}, \citenamefont
  {Eremeev}, \citenamefont {Koroteev}, \citenamefont {Kuznetsov}, \citenamefont
  {Freyse}, \citenamefont {S{\'a}nchez-Barriga}, \citenamefont {Amiraslanov},
  \citenamefont {Babanly}, \citenamefont {Mamedov}, \citenamefont {Abdullayev},
  \citenamefont {Zverev}, \citenamefont {Alfonsov}, \citenamefont {Kataev},
  \citenamefont {B{\"u}chner}, \citenamefont {Schwier}, \citenamefont {Kumar},
  \citenamefont {Kimura}, \citenamefont {Petaccia}, \citenamefont {Di~Santo},
  \citenamefont {Vidal}, \citenamefont {Schatz}, \citenamefont {Ki{\ss}ner},
  \citenamefont {{\"U}nzelmann}, \citenamefont {Min}, \citenamefont {Moser},
  \citenamefont {Peixoto}, \citenamefont {Reinert}, \citenamefont {Ernst},
  \citenamefont {Echenique}, \citenamefont {Isaeva},\ and\ \citenamefont
  {Chulkov}}]{otrokov-nat19}%
  \BibitemOpen
  \bibfield  {author} {\bibinfo {author} {\bibfnamefont {M.~M.}\ \bibnamefont
  {Otrokov}}, \bibinfo {author} {\bibfnamefont {I.~I.}\ \bibnamefont
  {Klimovskikh}}, \bibinfo {author} {\bibfnamefont {H.}~\bibnamefont
  {Bentmann}}, \bibinfo {author} {\bibfnamefont {D.}~\bibnamefont {Estyunin}},
  \bibinfo {author} {\bibfnamefont {A.}~\bibnamefont {Zeugner}}, \bibinfo
  {author} {\bibfnamefont {Z.~S.}\ \bibnamefont {Aliev}}, \bibinfo {author}
  {\bibfnamefont {S.}~\bibnamefont {Ga{\ss}}}, \bibinfo {author} {\bibfnamefont
  {A.~U.~B.}\ \bibnamefont {Wolter}}, \bibinfo {author} {\bibfnamefont {A.~V.}\
  \bibnamefont {Koroleva}}, \bibinfo {author} {\bibfnamefont {A.~M.}\
  \bibnamefont {Shikin}}, \bibinfo {author} {\bibfnamefont {M.}~\bibnamefont
  {Blanco-Rey}}, \bibinfo {author} {\bibfnamefont {M.}~\bibnamefont
  {Hoffmann}}, \bibinfo {author} {\bibfnamefont {I.~P.}\ \bibnamefont
  {Rusinov}}, \bibinfo {author} {\bibfnamefont {A.~Yu}\ \bibnamefont
  {Vyazovskaya}}, \bibinfo {author} {\bibfnamefont {S.~V.}\ \bibnamefont
  {Eremeev}}, \bibinfo {author} {\bibfnamefont {Yu~M.}\ \bibnamefont
  {Koroteev}}, \bibinfo {author} {\bibfnamefont {V.~M.}\ \bibnamefont
  {Kuznetsov}}, \bibinfo {author} {\bibfnamefont {F.}~\bibnamefont {Freyse}},
  \bibinfo {author} {\bibfnamefont {J.}~\bibnamefont {S{\'a}nchez-Barriga}},
  \bibinfo {author} {\bibfnamefont {I.~R.}\ \bibnamefont {Amiraslanov}},
  \bibinfo {author} {\bibfnamefont {M.~B.}\ \bibnamefont {Babanly}}, \bibinfo
  {author} {\bibfnamefont {N.~T.}\ \bibnamefont {Mamedov}}, \bibinfo {author}
  {\bibfnamefont {N.~A.}\ \bibnamefont {Abdullayev}}, \bibinfo {author}
  {\bibfnamefont {V.~N.}\ \bibnamefont {Zverev}}, \bibinfo {author}
  {\bibfnamefont {A.}~\bibnamefont {Alfonsov}}, \bibinfo {author}
  {\bibfnamefont {V.}~\bibnamefont {Kataev}}, \bibinfo {author} {\bibfnamefont
  {B.}~\bibnamefont {B{\"u}chner}}, \bibinfo {author} {\bibfnamefont {E.~F.}\
  \bibnamefont {Schwier}}, \bibinfo {author} {\bibfnamefont {S.}~\bibnamefont
  {Kumar}}, \bibinfo {author} {\bibfnamefont {A.}~\bibnamefont {Kimura}},
  \bibinfo {author} {\bibfnamefont {L.}~\bibnamefont {Petaccia}}, \bibinfo
  {author} {\bibfnamefont {G.}~\bibnamefont {Di~Santo}}, \bibinfo {author}
  {\bibfnamefont {R.~C.}\ \bibnamefont {Vidal}}, \bibinfo {author}
  {\bibfnamefont {S.}~\bibnamefont {Schatz}}, \bibinfo {author} {\bibfnamefont
  {K.}~\bibnamefont {Ki{\ss}ner}}, \bibinfo {author} {\bibfnamefont
  {M.}~\bibnamefont {{\"U}nzelmann}}, \bibinfo {author} {\bibfnamefont {C.~H.}\
  \bibnamefont {Min}}, \bibinfo {author} {\bibfnamefont {Simon}\ \bibnamefont
  {Moser}}, \bibinfo {author} {\bibfnamefont {T.~R.~F.}\ \bibnamefont
  {Peixoto}}, \bibinfo {author} {\bibfnamefont {F.}~\bibnamefont {Reinert}},
  \bibinfo {author} {\bibfnamefont {A.}~\bibnamefont {Ernst}}, \bibinfo
  {author} {\bibfnamefont {P.~M.}\ \bibnamefont {Echenique}}, \bibinfo {author}
  {\bibfnamefont {A.}~\bibnamefont {Isaeva}}, \ and\ \bibinfo {author}
  {\bibfnamefont {E.~V.}\ \bibnamefont {Chulkov}},\ }\bibfield  {title}
  {\enquote {\bibinfo {title} {Prediction and observation of an
  antiferromagnetic topological insulator},}\ }\href {\doibase
  10.1038/s41586-019-1840-9} {\bibfield  {journal} {\bibinfo  {journal}
  {Nature}\ }\textbf {\bibinfo {volume} {576}},\ \bibinfo {pages} {416--422}
  (\bibinfo {year} {2019})}\BibitemShut {NoStop}%
\bibitem [{\citenamefont {Li}\ \emph {et~al.}(2019{\natexlab{a}})\citenamefont
  {Li}, \citenamefont {Li}, \citenamefont {Du}, \citenamefont {Wang},
  \citenamefont {Gu}, \citenamefont {Zhang}, \citenamefont {He}, \citenamefont
  {Duan},\ and\ \citenamefont {Xu}}]{li-sc19}%
  \BibitemOpen
  \bibfield  {author} {\bibinfo {author} {\bibfnamefont {Jiaheng}\ \bibnamefont
  {Li}}, \bibinfo {author} {\bibfnamefont {Yang}\ \bibnamefont {Li}}, \bibinfo
  {author} {\bibfnamefont {Shiqiao}\ \bibnamefont {Du}}, \bibinfo {author}
  {\bibfnamefont {Zun}\ \bibnamefont {Wang}}, \bibinfo {author} {\bibfnamefont
  {Bing-Lin}\ \bibnamefont {Gu}}, \bibinfo {author} {\bibfnamefont
  {Shou-Cheng}\ \bibnamefont {Zhang}}, \bibinfo {author} {\bibfnamefont
  {Ke}~\bibnamefont {He}}, \bibinfo {author} {\bibfnamefont {Wenhui}\
  \bibnamefont {Duan}}, \ and\ \bibinfo {author} {\bibfnamefont {Yong}\
  \bibnamefont {Xu}},\ }\bibfield  {title} {\enquote {\bibinfo {title}
  {Intrinsic magnetic topological insulators in van der waals layered
  mnbi2te4-family materials},}\ }\href {\doibase 10.1126/sciadv.aaw5685}
  {\bibfield  {journal} {\bibinfo  {journal} {Science Advances}\ }\textbf
  {\bibinfo {volume} {5}} (\bibinfo {year} {2019}{\natexlab{a}}),\
  10.1126/sciadv.aaw5685}\BibitemShut {NoStop}%
\bibitem [{\citenamefont {Xu}\ \emph {et~al.}(2019)\citenamefont {Xu},
  \citenamefont {Song}, \citenamefont {Wang}, \citenamefont {Weng},\ and\
  \citenamefont {Dai}}]{xu-prl19}%
  \BibitemOpen
  \bibfield  {author} {\bibinfo {author} {\bibfnamefont {Yuanfeng}\
  \bibnamefont {Xu}}, \bibinfo {author} {\bibfnamefont {Zhida}\ \bibnamefont
  {Song}}, \bibinfo {author} {\bibfnamefont {Zhijun}\ \bibnamefont {Wang}},
  \bibinfo {author} {\bibfnamefont {Hongming}\ \bibnamefont {Weng}}, \ and\
  \bibinfo {author} {\bibfnamefont {Xi}~\bibnamefont {Dai}},\ }\bibfield
  {title} {\enquote {\bibinfo {title} {Higher-order topology of the axion
  insulator {EuIn$_2$As$_2$}},}\ }\href {\doibase
  10.1103/PhysRevLett.122.256402} {\bibfield  {journal} {\bibinfo  {journal}
  {Phys. Rev. Lett.}\ }\textbf {\bibinfo {volume} {122}},\ \bibinfo {pages}
  {256402} (\bibinfo {year} {2019})}\BibitemShut {NoStop}%
\bibitem [{\citenamefont {Li}\ \emph {et~al.}(2019{\natexlab{b}})\citenamefont
  {Li}, \citenamefont {Gao}, \citenamefont {Duan}, \citenamefont {Xu},
  \citenamefont {Zhu}, \citenamefont {Tian}, \citenamefont {Gao}, \citenamefont
  {Fan}, \citenamefont {Rao}, \citenamefont {Huang}, \citenamefont {Li},
  \citenamefont {Yan}, \citenamefont {Liu}, \citenamefont {Liu}, \citenamefont
  {Huang}, \citenamefont {Li}, \citenamefont {Liu}, \citenamefont {Zhang},
  \citenamefont {Zhang}, \citenamefont {Kondo}, \citenamefont {Shin},
  \citenamefont {Lei}, \citenamefont {Shi}, \citenamefont {Zhang},
  \citenamefont {Weng}, \citenamefont {Qian},\ and\ \citenamefont
  {Ding}}]{hang-prx19}%
  \BibitemOpen
  \bibfield  {author} {\bibinfo {author} {\bibfnamefont {Hang}\ \bibnamefont
  {Li}}, \bibinfo {author} {\bibfnamefont {Shun-Ye}\ \bibnamefont {Gao}},
  \bibinfo {author} {\bibfnamefont {Shao-Feng}\ \bibnamefont {Duan}}, \bibinfo
  {author} {\bibfnamefont {Yuan-Feng}\ \bibnamefont {Xu}}, \bibinfo {author}
  {\bibfnamefont {Ke-Jia}\ \bibnamefont {Zhu}}, \bibinfo {author}
  {\bibfnamefont {Shang-Jie}\ \bibnamefont {Tian}}, \bibinfo {author}
  {\bibfnamefont {Jia-Cheng}\ \bibnamefont {Gao}}, \bibinfo {author}
  {\bibfnamefont {Wen-Hui}\ \bibnamefont {Fan}}, \bibinfo {author}
  {\bibfnamefont {Zhi-Cheng}\ \bibnamefont {Rao}}, \bibinfo {author}
  {\bibfnamefont {Jie-Rui}\ \bibnamefont {Huang}}, \bibinfo {author}
  {\bibfnamefont {Jia-Jun}\ \bibnamefont {Li}}, \bibinfo {author}
  {\bibfnamefont {Da-Yu}\ \bibnamefont {Yan}}, \bibinfo {author} {\bibfnamefont
  {Zheng-Tai}\ \bibnamefont {Liu}}, \bibinfo {author} {\bibfnamefont
  {Wan-Ling}\ \bibnamefont {Liu}}, \bibinfo {author} {\bibfnamefont {Yao-Bo}\
  \bibnamefont {Huang}}, \bibinfo {author} {\bibfnamefont {Yu-Liang}\
  \bibnamefont {Li}}, \bibinfo {author} {\bibfnamefont {Yi}~\bibnamefont
  {Liu}}, \bibinfo {author} {\bibfnamefont {Guo-Bin}\ \bibnamefont {Zhang}},
  \bibinfo {author} {\bibfnamefont {Peng}\ \bibnamefont {Zhang}}, \bibinfo
  {author} {\bibfnamefont {Takeshi}\ \bibnamefont {Kondo}}, \bibinfo {author}
  {\bibfnamefont {Shik}\ \bibnamefont {Shin}}, \bibinfo {author} {\bibfnamefont
  {He-Chang}\ \bibnamefont {Lei}}, \bibinfo {author} {\bibfnamefont {You-Guo}\
  \bibnamefont {Shi}}, \bibinfo {author} {\bibfnamefont {Wen-Tao}\ \bibnamefont
  {Zhang}}, \bibinfo {author} {\bibfnamefont {Hong-Ming}\ \bibnamefont {Weng}},
  \bibinfo {author} {\bibfnamefont {Tian}\ \bibnamefont {Qian}}, \ and\
  \bibinfo {author} {\bibfnamefont {Hong}\ \bibnamefont {Ding}},\ }\bibfield
  {title} {\enquote {\bibinfo {title} {Dirac surface states in intrinsic
  magnetic topological insulators {${\mathrm{EuSn}}_{2}{\mathrm{As}}_{2}$} and
  {${\mathrm{MnBi}}_{2n}{\mathrm{Te}}_{3n+1}$}},}\ }\href {\doibase
  10.1103/PhysRevX.9.041039} {\bibfield  {journal} {\bibinfo  {journal} {Phys.
  Rev. X}\ }\textbf {\bibinfo {volume} {9}},\ \bibinfo {pages} {041039}
  (\bibinfo {year} {2019}{\natexlab{b}})}\BibitemShut {NoStop}%
\bibitem [{\citenamefont {Klimovskikh}\ \emph {et~al.}(2020)\citenamefont
  {Klimovskikh}, \citenamefont {Otrokov}, \citenamefont {Estyunin},
  \citenamefont {Eremeev}, \citenamefont {Filnov}, \citenamefont {Koroleva},
  \citenamefont {Shevchenko}, \citenamefont {Voroshnin}, \citenamefont
  {Rybkin}, \citenamefont {Rusinov}, \citenamefont {Blanco-Rey}, \citenamefont
  {Hoffmann}, \citenamefont {Aliev}, \citenamefont {Babanly}, \citenamefont
  {Amiraslanov}, \citenamefont {Abdullayev}, \citenamefont {Zverev},
  \citenamefont {Kimura}, \citenamefont {Tereshchenko}, \citenamefont {Kokh},
  \citenamefont {Petaccia}, \citenamefont {Di~Santo}, \citenamefont {Ernst},
  \citenamefont {Echenique}, \citenamefont {Mamedov}, \citenamefont {Shikin},\
  and\ \citenamefont {Chulkov}}]{klimovskikh-npj20}%
  \BibitemOpen
  \bibfield  {author} {\bibinfo {author} {\bibfnamefont {Ilya~I.}\ \bibnamefont
  {Klimovskikh}}, \bibinfo {author} {\bibfnamefont {Mikhail~M.}\ \bibnamefont
  {Otrokov}}, \bibinfo {author} {\bibfnamefont {Dmitry}\ \bibnamefont
  {Estyunin}}, \bibinfo {author} {\bibfnamefont {Sergey~V.}\ \bibnamefont
  {Eremeev}}, \bibinfo {author} {\bibfnamefont {Sergey~O.}\ \bibnamefont
  {Filnov}}, \bibinfo {author} {\bibfnamefont {Alexandra}\ \bibnamefont
  {Koroleva}}, \bibinfo {author} {\bibfnamefont {Eugene}\ \bibnamefont
  {Shevchenko}}, \bibinfo {author} {\bibfnamefont {Vladimir}\ \bibnamefont
  {Voroshnin}}, \bibinfo {author} {\bibfnamefont {Artem~G.}\ \bibnamefont
  {Rybkin}}, \bibinfo {author} {\bibfnamefont {Igor~P.}\ \bibnamefont
  {Rusinov}}, \bibinfo {author} {\bibfnamefont {Maria}\ \bibnamefont
  {Blanco-Rey}}, \bibinfo {author} {\bibfnamefont {Martin}\ \bibnamefont
  {Hoffmann}}, \bibinfo {author} {\bibfnamefont {Ziya~S.}\ \bibnamefont
  {Aliev}}, \bibinfo {author} {\bibfnamefont {Mahammad~B.}\ \bibnamefont
  {Babanly}}, \bibinfo {author} {\bibfnamefont {Imamaddin~R.}\ \bibnamefont
  {Amiraslanov}}, \bibinfo {author} {\bibfnamefont {Nadir~A.}\ \bibnamefont
  {Abdullayev}}, \bibinfo {author} {\bibfnamefont {Vladimir~N.}\ \bibnamefont
  {Zverev}}, \bibinfo {author} {\bibfnamefont {Akio}\ \bibnamefont {Kimura}},
  \bibinfo {author} {\bibfnamefont {Oleg~E.}\ \bibnamefont {Tereshchenko}},
  \bibinfo {author} {\bibfnamefont {Konstantin~A.}\ \bibnamefont {Kokh}},
  \bibinfo {author} {\bibfnamefont {Luca}\ \bibnamefont {Petaccia}}, \bibinfo
  {author} {\bibfnamefont {Giovanni}\ \bibnamefont {Di~Santo}}, \bibinfo
  {author} {\bibfnamefont {Arthur}\ \bibnamefont {Ernst}}, \bibinfo {author}
  {\bibfnamefont {Pedro~M.}\ \bibnamefont {Echenique}}, \bibinfo {author}
  {\bibfnamefont {Nazim~T.}\ \bibnamefont {Mamedov}}, \bibinfo {author}
  {\bibfnamefont {Alexander~M.}\ \bibnamefont {Shikin}}, \ and\ \bibinfo
  {author} {\bibfnamefont {Eugene~V.}\ \bibnamefont {Chulkov}},\ }\bibfield
  {title} {\enquote {\bibinfo {title} {Tunable 3d/2d magnetism in the
  (mnbi2te4)(bi2te3)m topological insulators family},}\ }\href {\doibase
  10.1038/s41535-020-00255-9} {\bibfield  {journal} {\bibinfo  {journal} {npj
  Quantum Materials}\ }\textbf {\bibinfo {volume} {5}},\ \bibinfo {pages} {54}
  (\bibinfo {year} {2020})}\BibitemShut {NoStop}%
\bibitem [{\citenamefont {Ma}\ \emph {et~al.}(2020)\citenamefont {Ma},
  \citenamefont {Wang}, \citenamefont {Nie}, \citenamefont {Yi}, \citenamefont
  {Xu}, \citenamefont {Li}, \citenamefont {Jandke}, \citenamefont {Wulfhekel},
  \citenamefont {Huang}, \citenamefont {West}, \citenamefont {Richard},
  \citenamefont {Chikina}, \citenamefont {Strocov}, \citenamefont {Mesot},
  \citenamefont {Weng}, \citenamefont {Zhang}, \citenamefont {Shi},
  \citenamefont {Qian}, \citenamefont {Shi},\ and\ \citenamefont
  {Ding}}]{ma-advmat20}%
  \BibitemOpen
  \bibfield  {author} {\bibinfo {author} {\bibfnamefont {Junzhang}\
  \bibnamefont {Ma}}, \bibinfo {author} {\bibfnamefont {Han}\ \bibnamefont
  {Wang}}, \bibinfo {author} {\bibfnamefont {Simin}\ \bibnamefont {Nie}},
  \bibinfo {author} {\bibfnamefont {Changjiang}\ \bibnamefont {Yi}}, \bibinfo
  {author} {\bibfnamefont {Yuanfeng}\ \bibnamefont {Xu}}, \bibinfo {author}
  {\bibfnamefont {Hang}\ \bibnamefont {Li}}, \bibinfo {author} {\bibfnamefont
  {Jasmin}\ \bibnamefont {Jandke}}, \bibinfo {author} {\bibfnamefont {Wulf}\
  \bibnamefont {Wulfhekel}}, \bibinfo {author} {\bibfnamefont {Yaobo}\
  \bibnamefont {Huang}}, \bibinfo {author} {\bibfnamefont {Damien}\
  \bibnamefont {West}}, \bibinfo {author} {\bibfnamefont {Pierre}\ \bibnamefont
  {Richard}}, \bibinfo {author} {\bibfnamefont {Alla}\ \bibnamefont {Chikina}},
  \bibinfo {author} {\bibfnamefont {Vladimir~N.}\ \bibnamefont {Strocov}},
  \bibinfo {author} {\bibfnamefont {Joël}\ \bibnamefont {Mesot}}, \bibinfo
  {author} {\bibfnamefont {Hongming}\ \bibnamefont {Weng}}, \bibinfo {author}
  {\bibfnamefont {Shengbai}\ \bibnamefont {Zhang}}, \bibinfo {author}
  {\bibfnamefont {Youguo}\ \bibnamefont {Shi}}, \bibinfo {author}
  {\bibfnamefont {Tian}\ \bibnamefont {Qian}}, \bibinfo {author} {\bibfnamefont
  {Ming}\ \bibnamefont {Shi}}, \ and\ \bibinfo {author} {\bibfnamefont {Hong}\
  \bibnamefont {Ding}},\ }\bibfield  {title} {\enquote {\bibinfo {title}
  {Emergence of nontrivial low-energy dirac fermions in antiferromagnetic
  eucd2as2},}\ }\href {\doibase https://doi.org/10.1002/adma.201907565}
  {\bibfield  {journal} {\bibinfo  {journal} {Advanced Materials}\ }\textbf
  {\bibinfo {volume} {32}},\ \bibinfo {pages} {1907565} (\bibinfo {year}
  {2020})},\ \Eprint
  {http://arxiv.org/abs/https://onlinelibrary.wiley.com/doi/pdf/10.1002/adma.201907565}
  {https://onlinelibrary.wiley.com/doi/pdf/10.1002/adma.201907565} \BibitemShut
  {NoStop}%
\bibitem [{\citenamefont {Xu}\ \emph {et~al.}(2020)\citenamefont {Xu},
  \citenamefont {Elcoro}, \citenamefont {Song}, \citenamefont {Wieder},
  \citenamefont {Vergniory}, \citenamefont {Regnault}, \citenamefont {Chen},
  \citenamefont {Felser},\ and\ \citenamefont {Bernevig}}]{xu-nat2020}%
  \BibitemOpen
  \bibfield  {author} {\bibinfo {author} {\bibfnamefont {Yuanfeng}\
  \bibnamefont {Xu}}, \bibinfo {author} {\bibfnamefont {Luis}\ \bibnamefont
  {Elcoro}}, \bibinfo {author} {\bibfnamefont {Zhi-Da}\ \bibnamefont {Song}},
  \bibinfo {author} {\bibfnamefont {Benjamin~J.}\ \bibnamefont {Wieder}},
  \bibinfo {author} {\bibfnamefont {M.~G.}\ \bibnamefont {Vergniory}}, \bibinfo
  {author} {\bibfnamefont {Nicolas}\ \bibnamefont {Regnault}}, \bibinfo
  {author} {\bibfnamefont {Yulin}\ \bibnamefont {Chen}}, \bibinfo {author}
  {\bibfnamefont {Claudia}\ \bibnamefont {Felser}}, \ and\ \bibinfo {author}
  {\bibfnamefont {B.~Andrei}\ \bibnamefont {Bernevig}},\ }\bibfield  {title}
  {\enquote {\bibinfo {title} {High-throughput calculations of magnetic
  topological materials},}\ }\href {\doibase 10.1038/s41586-020-2837-0}
  {\bibfield  {journal} {\bibinfo  {journal} {Nature}\ }\textbf {\bibinfo
  {volume} {586}},\ \bibinfo {pages} {702--707} (\bibinfo {year}
  {2020})}\BibitemShut {NoStop}%
\bibitem [{\citenamefont {Rosa}\ \emph {et~al.}(2020)\citenamefont {Rosa},
  \citenamefont {Xu}, \citenamefont {Rahn}, \citenamefont {Souza},
  \citenamefont {Kushwaha}, \citenamefont {Veiga}, \citenamefont {Bombardi},
  \citenamefont {Thomas}, \citenamefont {Janoschek}, \citenamefont {Bauer},
  \citenamefont {Chan}, \citenamefont {Wang}, \citenamefont {Thompson},
  \citenamefont {Harrison}, \citenamefont {Pagliuso}, \citenamefont
  {Bernevig},\ and\ \citenamefont {Ronning}}]{rosa-npjqm2020}%
  \BibitemOpen
  \bibfield  {author} {\bibinfo {author} {\bibfnamefont {Priscila}\
  \bibnamefont {Rosa}}, \bibinfo {author} {\bibfnamefont {Yuanfeng}\
  \bibnamefont {Xu}}, \bibinfo {author} {\bibfnamefont {Marein}\ \bibnamefont
  {Rahn}}, \bibinfo {author} {\bibfnamefont {Jean}\ \bibnamefont {Souza}},
  \bibinfo {author} {\bibfnamefont {Satya}\ \bibnamefont {Kushwaha}}, \bibinfo
  {author} {\bibfnamefont {Larissa}\ \bibnamefont {Veiga}}, \bibinfo {author}
  {\bibfnamefont {Alessandro}\ \bibnamefont {Bombardi}}, \bibinfo {author}
  {\bibfnamefont {Sean}\ \bibnamefont {Thomas}}, \bibinfo {author}
  {\bibfnamefont {Marc}\ \bibnamefont {Janoschek}}, \bibinfo {author}
  {\bibfnamefont {Eric}\ \bibnamefont {Bauer}}, \bibinfo {author}
  {\bibfnamefont {Mun}\ \bibnamefont {Chan}}, \bibinfo {author} {\bibfnamefont
  {Zhijun}\ \bibnamefont {Wang}}, \bibinfo {author} {\bibfnamefont {Joe}\
  \bibnamefont {Thompson}}, \bibinfo {author} {\bibfnamefont {Neil}\
  \bibnamefont {Harrison}}, \bibinfo {author} {\bibfnamefont {Pascoal}\
  \bibnamefont {Pagliuso}}, \bibinfo {author} {\bibfnamefont {Andrei}\
  \bibnamefont {Bernevig}}, \ and\ \bibinfo {author} {\bibfnamefont {Filip}\
  \bibnamefont {Ronning}},\ }\bibfield  {title} {\enquote {\bibinfo {title}
  {Colossal magnetoresistance in a nonsymmorphic antiferromagnetic
  insulator},}\ }\href {\doibase 10.1038/s41535-020-00256-8} {\bibfield
  {journal} {\bibinfo  {journal} {npj Quantum Materials}\ }\textbf {\bibinfo
  {volume} {5}},\ \bibinfo {pages} {52} (\bibinfo {year} {2020})}\BibitemShut
  {NoStop}%
\bibitem [{\citenamefont {Pierantozzi}\ \emph {et~al.}(2022)\citenamefont
  {Pierantozzi}, \citenamefont {De~Vita}, \citenamefont {Bigi}, \citenamefont
  {Gui}, \citenamefont {Tien}, \citenamefont {Mondal}, \citenamefont {Mazzola},
  \citenamefont {Fujii}, \citenamefont {Vobornik}, \citenamefont {Vinai},
  \citenamefont {Sala}, \citenamefont {Africh}, \citenamefont {Lee},
  \citenamefont {Rossi}, \citenamefont {Chang}, \citenamefont {Xie},
  \citenamefont {Cava},\ and\ \citenamefont
  {Panaccione}}]{pierantozzie-pnas22}%
  \BibitemOpen
  \bibfield  {author} {\bibinfo {author} {\bibfnamefont {Gian~Marco}\
  \bibnamefont {Pierantozzi}}, \bibinfo {author} {\bibfnamefont {Alessandro}\
  \bibnamefont {De~Vita}}, \bibinfo {author} {\bibfnamefont {Chiara}\
  \bibnamefont {Bigi}}, \bibinfo {author} {\bibfnamefont {Xin}\ \bibnamefont
  {Gui}}, \bibinfo {author} {\bibfnamefont {Hung-Ju}\ \bibnamefont {Tien}},
  \bibinfo {author} {\bibfnamefont {Debashis}\ \bibnamefont {Mondal}}, \bibinfo
  {author} {\bibfnamefont {Federico}\ \bibnamefont {Mazzola}}, \bibinfo
  {author} {\bibfnamefont {Jun}\ \bibnamefont {Fujii}}, \bibinfo {author}
  {\bibfnamefont {Ivana}\ \bibnamefont {Vobornik}}, \bibinfo {author}
  {\bibfnamefont {Giovanni}\ \bibnamefont {Vinai}}, \bibinfo {author}
  {\bibfnamefont {Alessandro}\ \bibnamefont {Sala}}, \bibinfo {author}
  {\bibfnamefont {Cristina}\ \bibnamefont {Africh}}, \bibinfo {author}
  {\bibfnamefont {Tien-Lin}\ \bibnamefont {Lee}}, \bibinfo {author}
  {\bibfnamefont {Giorgio}\ \bibnamefont {Rossi}}, \bibinfo {author}
  {\bibfnamefont {Tay-Rong}\ \bibnamefont {Chang}}, \bibinfo {author}
  {\bibfnamefont {Weiwei}\ \bibnamefont {Xie}}, \bibinfo {author}
  {\bibfnamefont {Robert~J.}\ \bibnamefont {Cava}}, \ and\ \bibinfo {author}
  {\bibfnamefont {Giancarlo}\ \bibnamefont {Panaccione}},\ }\bibfield  {title}
  {\enquote {\bibinfo {title} {Evidence of magnetism-induced topological
  protection in the axion insulator candidate {EuSn$_2$P$_2$}},}\ }\href
  {\doibase 10.1073/pnas.2116575119} {\bibfield  {journal} {\bibinfo  {journal}
  {Proceedings of the National Academy of Sciences}\ }\textbf {\bibinfo
  {volume} {119}} (\bibinfo {year} {2022}),\ 10.1073/pnas.2116575119},\ \Eprint
  {http://arxiv.org/abs/https://www.pnas.org/content/119/4/e2116575119.full.pdf}
  {https://www.pnas.org/content/119/4/e2116575119.full.pdf} \BibitemShut
  {NoStop}%
\bibitem [{\citenamefont {Deng}\ \emph {et~al.}(2020)\citenamefont {Deng},
  \citenamefont {Yu}, \citenamefont {Shi}, \citenamefont {Guo}, \citenamefont
  {Xu}, \citenamefont {Wang}, \citenamefont {Chen},\ and\ \citenamefont
  {Zhang}}]{deng-sc20}%
  \BibitemOpen
  \bibfield  {author} {\bibinfo {author} {\bibfnamefont {Yujun}\ \bibnamefont
  {Deng}}, \bibinfo {author} {\bibfnamefont {Yijun}\ \bibnamefont {Yu}},
  \bibinfo {author} {\bibfnamefont {Meng~Zhu}\ \bibnamefont {Shi}}, \bibinfo
  {author} {\bibfnamefont {Zhongxun}\ \bibnamefont {Guo}}, \bibinfo {author}
  {\bibfnamefont {Zihan}\ \bibnamefont {Xu}}, \bibinfo {author} {\bibfnamefont
  {Jing}\ \bibnamefont {Wang}}, \bibinfo {author} {\bibfnamefont {Xian~Hui}\
  \bibnamefont {Chen}}, \ and\ \bibinfo {author} {\bibfnamefont {Yuanbo}\
  \bibnamefont {Zhang}},\ }\bibfield  {title} {\enquote {\bibinfo {title}
  {Quantum anomalous hall effect in intrinsic magnetic topological insulator
  {MnBi}$_{2}${Te}$_4$},}\ }\href {\doibase 10.1126/science.aax8156} {\bibfield
   {journal} {\bibinfo  {journal} {Science}\ }\textbf {\bibinfo {volume}
  {367}},\ \bibinfo {pages} {895--900} (\bibinfo {year} {2020})}\BibitemShut
  {NoStop}%
\bibitem [{\citenamefont {Liu}\ \emph {et~al.}(2020)\citenamefont {Liu},
  \citenamefont {Wang}, \citenamefont {Li}, \citenamefont {Wu}, \citenamefont
  {Li}, \citenamefont {Li}, \citenamefont {He}, \citenamefont {Xu},
  \citenamefont {Zhang},\ and\ \citenamefont {Wang}}]{liu-natmat20}%
  \BibitemOpen
  \bibfield  {author} {\bibinfo {author} {\bibfnamefont {Chang}\ \bibnamefont
  {Liu}}, \bibinfo {author} {\bibfnamefont {Yongchao}\ \bibnamefont {Wang}},
  \bibinfo {author} {\bibfnamefont {Hao}\ \bibnamefont {Li}}, \bibinfo {author}
  {\bibfnamefont {Yang}\ \bibnamefont {Wu}}, \bibinfo {author} {\bibfnamefont
  {Yaoxin}\ \bibnamefont {Li}}, \bibinfo {author} {\bibfnamefont {Jiaheng}\
  \bibnamefont {Li}}, \bibinfo {author} {\bibfnamefont {Ke}~\bibnamefont {He}},
  \bibinfo {author} {\bibfnamefont {Yong}\ \bibnamefont {Xu}}, \bibinfo
  {author} {\bibfnamefont {Jinsong}\ \bibnamefont {Zhang}}, \ and\ \bibinfo
  {author} {\bibfnamefont {Yayu}\ \bibnamefont {Wang}},\ }\bibfield  {title}
  {\enquote {\bibinfo {title} {Robust axion insulator and chern insulator
  phases in a two-dimensional antiferromagnetic topological insulator},}\
  }\href {\doibase 10.1038/s41563-019-0573-3} {\bibfield  {journal} {\bibinfo
  {journal} {Nature Materials}\ }\textbf {\bibinfo {volume} {19}},\ \bibinfo
  {pages} {522--527} (\bibinfo {year} {2020})}\BibitemShut {NoStop}%
\bibitem [{\citenamefont {Wang}\ \emph
  {et~al.}(2021{\natexlab{b}})\citenamefont {Wang}, \citenamefont {Ge},
  \citenamefont {Li}, \citenamefont {Liu}, \citenamefont {Xu},\ and\
  \citenamefont {Wang}}]{wang21-scdir}%
  \BibitemOpen
  \bibfield  {author} {\bibinfo {author} {\bibfnamefont {Pinyuan}\ \bibnamefont
  {Wang}}, \bibinfo {author} {\bibfnamefont {Jun}\ \bibnamefont {Ge}}, \bibinfo
  {author} {\bibfnamefont {Jiaheng}\ \bibnamefont {Li}}, \bibinfo {author}
  {\bibfnamefont {Yanzhao}\ \bibnamefont {Liu}}, \bibinfo {author}
  {\bibfnamefont {Yong}\ \bibnamefont {Xu}}, \ and\ \bibinfo {author}
  {\bibfnamefont {Jian}\ \bibnamefont {Wang}},\ }\bibfield  {title} {\enquote
  {\bibinfo {title} {Intrinsic magnetic topological insulators},}\ }\href
  {\doibase https://doi.org/10.1016/j.xinn.2021.100098} {\bibfield  {journal}
  {\bibinfo  {journal} {The Innovation}\ }\textbf {\bibinfo {volume} {2}},\
  \bibinfo {pages} {100098} (\bibinfo {year} {2021}{\natexlab{b}})}\BibitemShut
  {NoStop}%
\bibitem [{\citenamefont {Kauzlarich}\ \emph {et~al.}(2017)\citenamefont
  {Kauzlarich}, \citenamefont {Zevalkink}, \citenamefont {Toberer},\ and\
  \citenamefont {Snyder}}]{kauzlarich-tmd17}%
  \BibitemOpen
  \bibfield  {author} {\bibinfo {author} {\bibfnamefont {Susan~M.}\
  \bibnamefont {Kauzlarich}}, \bibinfo {author} {\bibfnamefont {Alex}\
  \bibnamefont {Zevalkink}}, \bibinfo {author} {\bibfnamefont {Eric}\
  \bibnamefont {Toberer}}, \ and\ \bibinfo {author} {\bibfnamefont {G.~Jeff}\
  \bibnamefont {Snyder}},\ }\bibfield  {title} {\enquote {\bibinfo {title}
  {Chapter 1 zintl phases: Recent developments in thermoelectrics and future
  outlook},}\ }in\ \href {\doibase 10.1039/9781782624042-00001} {\emph
  {\bibinfo {booktitle} {Thermoelectric Materials and Devices}}}\ (\bibinfo
  {publisher} {The Royal Society of Chemistry},\ \bibinfo {year} {2017})\ pp.\
  \bibinfo {pages} {1--26}\BibitemShut {NoStop}%
\bibitem [{\citenamefont {Deakin}\ \emph {et~al.}(2002)\citenamefont {Deakin},
  \citenamefont {Lam}, \citenamefont {Marsiglio},\ and\ \citenamefont
  {Mar}}]{deakin-jac02}%
  \BibitemOpen
  \bibfield  {author} {\bibinfo {author} {\bibfnamefont {Laura}\ \bibnamefont
  {Deakin}}, \bibinfo {author} {\bibfnamefont {Robert}\ \bibnamefont {Lam}},
  \bibinfo {author} {\bibfnamefont {Frank}\ \bibnamefont {Marsiglio}}, \ and\
  \bibinfo {author} {\bibfnamefont {Arthur}\ \bibnamefont {Mar}},\ }\bibfield
  {title} {\enquote {\bibinfo {title} {Superconductivity in {Ba2Sn3Sb6} and
  {SrSn3Sb4}},}\ }\href {\doibase
  https://doi.org/10.1016/S0925-8388(02)00216-5} {\bibfield  {journal}
  {\bibinfo  {journal} {Journal of Alloys and Compounds}\ }\textbf {\bibinfo
  {volume} {338}},\ \bibinfo {pages} {69--72} (\bibinfo {year} {2002})},\
  \bibinfo {note} {special Issue to Honor Professor H. Fritz
  Franzen}\BibitemShut {NoStop}%
\bibitem [{\citenamefont {Chan}\ \emph {et~al.}(1997)\citenamefont {Chan},
  \citenamefont {Kauzlarich}, \citenamefont {Klavins}, \citenamefont
  {Shelton},\ and\ \citenamefont {Webb}}]{chan-cm97}%
  \BibitemOpen
  \bibfield  {author} {\bibinfo {author} {\bibfnamefont {Julia~Y.}\
  \bibnamefont {Chan}}, \bibinfo {author} {\bibfnamefont {Susan~M.}\
  \bibnamefont {Kauzlarich}}, \bibinfo {author} {\bibfnamefont {Peter}\
  \bibnamefont {Klavins}}, \bibinfo {author} {\bibfnamefont {Robert~N.}\
  \bibnamefont {Shelton}}, \ and\ \bibinfo {author} {\bibfnamefont {David~J.}\
  \bibnamefont {Webb}},\ }\bibfield  {title} {\enquote {\bibinfo {title}
  {Colossal magnetoresistance in the transition-metal zintl compound
  {Eu$_{14}$MnSb$_{11}$}},}\ }\href {\doibase 10.1021/cm9704241} {\bibfield
  {journal} {\bibinfo  {journal} {Chemistry of Materials}\ }\textbf {\bibinfo
  {volume} {9}},\ \bibinfo {pages} {3132--3135} (\bibinfo {year}
  {1997})}\BibitemShut {NoStop}%
\bibitem [{\citenamefont {Liu}\ and\ \citenamefont {Xia}(2019)}]{liu-jssc19}%
  \BibitemOpen
  \bibfield  {author} {\bibinfo {author} {\bibfnamefont {Ke-Feng}\ \bibnamefont
  {Liu}}\ and\ \bibinfo {author} {\bibfnamefont {Sheng-Qing}\ \bibnamefont
  {Xia}},\ }\bibfield  {title} {\enquote {\bibinfo {title} {Recent progresses
  on thermoelectric zintl phases: Structures, materials and optimization},}\
  }\href {\doibase https://doi.org/10.1016/j.jssc.2018.11.030} {\bibfield
  {journal} {\bibinfo  {journal} {Journal of Solid State Chemistry}\ }\textbf
  {\bibinfo {volume} {270}},\ \bibinfo {pages} {252--264} (\bibinfo {year}
  {2019})}\BibitemShut {NoStop}%
\bibitem [{\citenamefont {Park}\ \emph {et~al.}(2002)\citenamefont {Park},
  \citenamefont {Choi}, \citenamefont {Kang},\ and\ \citenamefont
  {Kim}}]{park-jmc02}%
  \BibitemOpen
  \bibfield  {author} {\bibinfo {author} {\bibfnamefont {Seon-Mi}\ \bibnamefont
  {Park}}, \bibinfo {author} {\bibfnamefont {Eun~Sang}\ \bibnamefont {Choi}},
  \bibinfo {author} {\bibfnamefont {Woun}\ \bibnamefont {Kang}}, \ and\
  \bibinfo {author} {\bibfnamefont {Sung-Jin}\ \bibnamefont {Kim}},\ }\bibfield
   {title} {\enquote {\bibinfo {title} {{Eu$_5$In$_2$Sb$_6$},
  {Eu$_5$In$_{2-x}$Zn$_x$Sb$_6$}: rare earth zintl phases with narrow band
  gaps},}\ }\href {\doibase 10.1039/B106812A} {\bibfield  {journal} {\bibinfo
  {journal} {J. Mater. Chem.}\ }\textbf {\bibinfo {volume} {12}},\ \bibinfo
  {pages} {1839--1843} (\bibinfo {year} {2002})}\BibitemShut {NoStop}%
\bibitem [{\citenamefont {Fu}\ and\ \citenamefont {Kane}(2007)}]{fu-prb07}%
  \BibitemOpen
  \bibfield  {author} {\bibinfo {author} {\bibfnamefont {L.}~\bibnamefont
  {Fu}}\ and\ \bibinfo {author} {\bibfnamefont {C.~L.}\ \bibnamefont {Kane}},\
  }\bibfield  {title} {\enquote {\bibinfo {title} {Topological insulators with
  inversion symmetry},}\ }\href {\doibase 10.1103/PhysRevB.76.045302}
  {\bibfield  {journal} {\bibinfo  {journal} {Phys. Rev. B}\ }\textbf {\bibinfo
  {volume} {76}},\ \bibinfo {pages} {045302} (\bibinfo {year}
  {2007})}\BibitemShut {NoStop}%
\bibitem [{\citenamefont {Bradlyn}\ \emph {et~al.}(2017)\citenamefont
  {Bradlyn}, \citenamefont {Elcoro}, \citenamefont {Cano}, \citenamefont
  {Vergniory}, \citenamefont {Wang}, \citenamefont {Felser}, \citenamefont
  {Aroyo},\ and\ \citenamefont {Bernevig}}]{bradlyn-nat17}%
  \BibitemOpen
  \bibfield  {author} {\bibinfo {author} {\bibfnamefont {B.}~\bibnamefont
  {Bradlyn}}, \bibinfo {author} {\bibfnamefont {L.}~\bibnamefont {Elcoro}},
  \bibinfo {author} {\bibfnamefont {J.}~\bibnamefont {Cano}}, \bibinfo {author}
  {\bibfnamefont {M.~G.}\ \bibnamefont {Vergniory}}, \bibinfo {author}
  {\bibfnamefont {Z.}~\bibnamefont {Wang}}, \bibinfo {author} {\bibfnamefont
  {C.}~\bibnamefont {Felser}}, \bibinfo {author} {\bibfnamefont {M.~I.}\
  \bibnamefont {Aroyo}}, \ and\ \bibinfo {author} {\bibfnamefont {B.~A.}\
  \bibnamefont {Bernevig}},\ }\bibfield  {title} {\enquote {\bibinfo {title}
  {Topological quantum chemistry},}\ }\href {\doibase 10.1038/nature23268}
  {\bibfield  {journal} {\bibinfo  {journal} {Nature}\ }\textbf {\bibinfo
  {volume} {547}},\ \bibinfo {pages} {298} (\bibinfo {year}
  {2017})}\BibitemShut {NoStop}%
\bibitem [{\citenamefont {Vergniory}\ \emph {et~al.}(2019)\citenamefont
  {Vergniory}, \citenamefont {Elcoro}, \citenamefont {Felser}, \citenamefont
  {Regnault}, \citenamefont {Bernevig},\ and\ \citenamefont
  {Wang}}]{vergniory-nat19}%
  \BibitemOpen
  \bibfield  {author} {\bibinfo {author} {\bibfnamefont {M.~G.}\ \bibnamefont
  {Vergniory}}, \bibinfo {author} {\bibfnamefont {L.}~\bibnamefont {Elcoro}},
  \bibinfo {author} {\bibfnamefont {Claudia}\ \bibnamefont {Felser}}, \bibinfo
  {author} {\bibfnamefont {Nicolas}\ \bibnamefont {Regnault}}, \bibinfo
  {author} {\bibfnamefont {B.~Andrei}\ \bibnamefont {Bernevig}}, \ and\
  \bibinfo {author} {\bibfnamefont {Zhijun}\ \bibnamefont {Wang}},\ }\bibfield
  {title} {\enquote {\bibinfo {title} {A complete catalogue of high-quality
  topological materials},}\ }\href {\doibase 10.1038/s41586-019-0954-4}
  {\bibfield  {journal} {\bibinfo  {journal} {Nature}\ }\textbf {\bibinfo
  {volume} {566}},\ \bibinfo {pages} {480--485} (\bibinfo {year}
  {2019})}\BibitemShut {NoStop}%
\bibitem [{\citenamefont {Vergniory}\ \emph {et~al.}(2021)\citenamefont
  {Vergniory}, \citenamefont {Wieder}, \citenamefont {Elcoro}, \citenamefont
  {Parkin}, \citenamefont {Felser}, \citenamefont {Bernevig},\ and\
  \citenamefont {Regnault}}]{vergniory-arxiv21}%
  \BibitemOpen
  \bibfield  {author} {\bibinfo {author} {\bibfnamefont {Maia~G.}\ \bibnamefont
  {Vergniory}}, \bibinfo {author} {\bibfnamefont {Benjamin~J.}\ \bibnamefont
  {Wieder}}, \bibinfo {author} {\bibfnamefont {Luis}\ \bibnamefont {Elcoro}},
  \bibinfo {author} {\bibfnamefont {Stuart S.~P.}\ \bibnamefont {Parkin}},
  \bibinfo {author} {\bibfnamefont {Claudia}\ \bibnamefont {Felser}}, \bibinfo
  {author} {\bibfnamefont {B.~Andrei}\ \bibnamefont {Bernevig}}, \ and\
  \bibinfo {author} {\bibfnamefont {Nicolas}\ \bibnamefont {Regnault}},\
  }\href@noop {} {\enquote {\bibinfo {title} {All topological bands of all
  stoichiometric materials},}\ } (\bibinfo {year} {2021}),\ \Eprint
  {http://arxiv.org/abs/arXiv:2105.09954} {arXiv:2105.09954} \BibitemShut
  {NoStop}%
\bibitem [{\citenamefont {Aroyo}\ \emph
  {et~al.}(2006{\natexlab{a}})\citenamefont {Aroyo}, \citenamefont {Kirov},
  \citenamefont {Capillas}, \citenamefont {Perez-Mato},\ and\ \citenamefont
  {Wondratschek}}]{aroyo-acs06}%
  \BibitemOpen
  \bibfield  {author} {\bibinfo {author} {\bibfnamefont {Mois~I.}\ \bibnamefont
  {Aroyo}}, \bibinfo {author} {\bibfnamefont {Asen}\ \bibnamefont {Kirov}},
  \bibinfo {author} {\bibfnamefont {Cesar}\ \bibnamefont {Capillas}}, \bibinfo
  {author} {\bibfnamefont {J.~M.}\ \bibnamefont {Perez-Mato}}, \ and\ \bibinfo
  {author} {\bibfnamefont {Hans}\ \bibnamefont {Wondratschek}},\ }\bibfield
  {title} {\enquote {\bibinfo {title} {{Bilbao Crystallographic Server. II.
  Representations of crystallographic point groups and space groups}},}\ }\href
  {\doibase 10.1107/S0108767305040286} {\bibfield  {journal} {\bibinfo
  {journal} {Acta Crystallographica Section A}\ }\textbf {\bibinfo {volume}
  {62}},\ \bibinfo {pages} {115--128} (\bibinfo {year}
  {2006}{\natexlab{a}})}\BibitemShut {NoStop}%
\bibitem [{\citenamefont {Aroyo}\ \emph
  {et~al.}(2006{\natexlab{b}})\citenamefont {Aroyo}, \citenamefont
  {Perez-Mato}, \citenamefont {Capillas}, \citenamefont {Kroumova},
  \citenamefont {Ivantchev}, \citenamefont {Madariaga}, \citenamefont {Kirov},\
  and\ \citenamefont {Wondratschek}}]{aroyo-zfk06}%
  \BibitemOpen
  \bibfield  {author} {\bibinfo {author} {\bibfnamefont {Mois~Ilia}\
  \bibnamefont {Aroyo}}, \bibinfo {author} {\bibfnamefont {Juan~Manuel}\
  \bibnamefont {Perez-Mato}}, \bibinfo {author} {\bibfnamefont {Cesar}\
  \bibnamefont {Capillas}}, \bibinfo {author} {\bibfnamefont {Eli}\
  \bibnamefont {Kroumova}}, \bibinfo {author} {\bibfnamefont {Svetoslav}\
  \bibnamefont {Ivantchev}}, \bibinfo {author} {\bibfnamefont {Gotzon}\
  \bibnamefont {Madariaga}}, \bibinfo {author} {\bibfnamefont {Asen}\
  \bibnamefont {Kirov}}, \ and\ \bibinfo {author} {\bibfnamefont {Hans}\
  \bibnamefont {Wondratschek}},\ }\bibfield  {title} {\enquote {\bibinfo
  {title} {Bilbao crystallographic server: I. databases and crystallographic
  computing programs},}\ }\href {\doibase doi:10.1524/zkri.2006.221.1.15}
  {\bibfield  {journal} {\bibinfo  {journal} {Zeitschrift für Kristallographie
  - Crystalline Materials}\ }\textbf {\bibinfo {volume} {221}},\ \bibinfo
  {pages} {15--27} (\bibinfo {year} {2006}{\natexlab{b}})}\BibitemShut
  {NoStop}%
\bibitem [{\citenamefont {Aroyo}\ \emph {et~al.}(2011)\citenamefont {Aroyo},
  \citenamefont {Perez-Mato}, \citenamefont {Orobengoa}, \citenamefont {Tasci},
  \citenamefont {De~la Flor~Martin},\ and\ \citenamefont
  {Kirov}}]{aroyo-bcc11}%
  \BibitemOpen
  \bibfield  {author} {\bibinfo {author} {\bibfnamefont {M.~I.}\ \bibnamefont
  {Aroyo}}, \bibinfo {author} {\bibfnamefont {J.}~\bibnamefont {Perez-Mato}},
  \bibinfo {author} {\bibfnamefont {D}~\bibnamefont {Orobengoa}}, \bibinfo
  {author} {\bibfnamefont {Emre}\ \bibnamefont {Tasci}}, \bibinfo {author}
  {\bibfnamefont {Gemma}\ \bibnamefont {De~la Flor~Martin}}, \ and\ \bibinfo
  {author} {\bibfnamefont {A}~\bibnamefont {Kirov}},\ }\bibfield  {title}
  {\enquote {\bibinfo {title} {Crystallography online: Bilbao crystallographic
  server},}\ }\href@noop {} {\bibfield  {journal} {\bibinfo  {journal}
  {Bulgarian Chemical Communications}\ }\textbf {\bibinfo {volume} {43}},\
  \bibinfo {pages} {183--197} (\bibinfo {year} {2011})}\BibitemShut {NoStop}%
\bibitem [{Note1()}]{Note1}%
  \BibitemOpen
  \bibinfo {note} {$y_{\protect \rm In}=0.2419$ was used in their calculation
  while $y_{\protect \rm In}=0.2149$ is reported in \protect \citet
  {park-jmc02}.}\BibitemShut {Stop}%
\bibitem [{\citenamefont {Childs}\ \emph {et~al.}(2019)\citenamefont {Childs},
  \citenamefont {Baranets},\ and\ \citenamefont {Bobev}}]{childs-jssc19}%
  \BibitemOpen
  \bibfield  {author} {\bibinfo {author} {\bibfnamefont {Amanda~B.}\
  \bibnamefont {Childs}}, \bibinfo {author} {\bibfnamefont {Sviatoslav}\
  \bibnamefont {Baranets}}, \ and\ \bibinfo {author} {\bibfnamefont {Svilen}\
  \bibnamefont {Bobev}},\ }\bibfield  {title} {\enquote {\bibinfo {title} {Five
  new ternary indium-arsenides discovered. synthesis and structural
  characterization of the zintl phases {Sr$_3$In$_2$As$_4$, Ba$_3$In$_2$As$_4$,
  Eu$_3$In$_2$As$_4$, Sr$_5$In$_2$As$_6$ and Eu$_5$In$_2$As$_6$}},}\ }\href
  {\doibase https://doi.org/10.1016/j.jssc.2019.07.050} {\bibfield  {journal}
  {\bibinfo  {journal} {Journal of Solid State Chemistry}\ }\textbf {\bibinfo
  {volume} {278}},\ \bibinfo {pages} {120889} (\bibinfo {year}
  {2019})}\BibitemShut {NoStop}%
\bibitem [{\citenamefont {Radzieowski}\ \emph {et~al.}(2020)\citenamefont
  {Radzieowski}, \citenamefont {Stegemann}, \citenamefont {Klenner},
  \citenamefont {Zhang}, \citenamefont {Fokwa},\ and\ \citenamefont
  {Janka}}]{radzieowski-mcf20}%
  \BibitemOpen
  \bibfield  {author} {\bibinfo {author} {\bibfnamefont {Mathis}\ \bibnamefont
  {Radzieowski}}, \bibinfo {author} {\bibfnamefont {Frank}\ \bibnamefont
  {Stegemann}}, \bibinfo {author} {\bibfnamefont {Steffen}\ \bibnamefont
  {Klenner}}, \bibinfo {author} {\bibfnamefont {Yuemei}\ \bibnamefont {Zhang}},
  \bibinfo {author} {\bibfnamefont {Boniface P.~T.}\ \bibnamefont {Fokwa}}, \
  and\ \bibinfo {author} {\bibfnamefont {Oliver}\ \bibnamefont {Janka}},\
  }\bibfield  {title} {\enquote {\bibinfo {title} {On the divalent character of
  the eu atoms in the ternary zintl phases {Eu$_5$In$_2$Pn$_6$} and
  {Eu$_3$MAs$_3$} {(Pn = As–Bi; M = Al,Ga)}},}\ }\href {\doibase
  10.1039/C9QM00703B} {\bibfield  {journal} {\bibinfo  {journal} {Mater. Chem.
  Front.}\ }\textbf {\bibinfo {volume} {4}},\ \bibinfo {pages} {1231--1248}
  (\bibinfo {year} {2020})}\BibitemShut {NoStop}%
\bibitem [{\citenamefont {Subbarao}\ \emph {et~al.}(2016)\citenamefont
  {Subbarao}, \citenamefont {Sarkar}, \citenamefont {Joseph},\ and\
  \citenamefont {Peter}}]{subbarao-jac16}%
  \BibitemOpen
  \bibfield  {author} {\bibinfo {author} {\bibfnamefont {Udumula}\ \bibnamefont
  {Subbarao}}, \bibinfo {author} {\bibfnamefont {Sumanta}\ \bibnamefont
  {Sarkar}}, \bibinfo {author} {\bibfnamefont {Boby}\ \bibnamefont {Joseph}}, \
  and\ \bibinfo {author} {\bibfnamefont {Sebastian~C.}\ \bibnamefont {Peter}},\
  }\bibfield  {title} {\enquote {\bibinfo {title} {Magnetic and {X-ray}
  absorption studies on the {RE$_5$X$_2$Sb$_6$ (RE=Eu, Yb; X=Al, Ga, In)}
  compounds},}\ }\href {\doibase https://doi.org/10.1016/j.jallcom.2015.10.232}
  {\bibfield  {journal} {\bibinfo  {journal} {Journal of Alloys and Compounds}\
  }\textbf {\bibinfo {volume} {658}},\ \bibinfo {pages} {395--401} (\bibinfo
  {year} {2016})}\BibitemShut {NoStop}%
\bibitem [{\citenamefont {Teo}\ \emph {et~al.}(2008)\citenamefont {Teo},
  \citenamefont {Fu},\ and\ \citenamefont {Kane}}]{teo-prb08}%
  \BibitemOpen
  \bibfield  {author} {\bibinfo {author} {\bibfnamefont {Jeffrey C.~Y.}\
  \bibnamefont {Teo}}, \bibinfo {author} {\bibfnamefont {Liang}\ \bibnamefont
  {Fu}}, \ and\ \bibinfo {author} {\bibfnamefont {C.~L.}\ \bibnamefont
  {Kane}},\ }\bibfield  {title} {\enquote {\bibinfo {title} {Surface states and
  topological invariants in three-dimensional topological insulators:
  Application to {${\text{Bi}}_{1\ensuremath{-}x}{\text{Sb}}_{x}$}},}\ }\href
  {\doibase 10.1103/PhysRevB.78.045426} {\bibfield  {journal} {\bibinfo
  {journal} {Phys. Rev. B}\ }\textbf {\bibinfo {volume} {78}},\ \bibinfo
  {pages} {045426} (\bibinfo {year} {2008})}\BibitemShut {NoStop}%
\bibitem [{\citenamefont {Hsieh}\ \emph {et~al.}(2008)\citenamefont {Hsieh},
  \citenamefont {Qian}, \citenamefont {Wray}, \citenamefont {Xia},
  \citenamefont {Hor}, \citenamefont {Cava},\ and\ \citenamefont
  {Hasan}}]{hsieh-nat08}%
  \BibitemOpen
  \bibfield  {author} {\bibinfo {author} {\bibfnamefont {D.}~\bibnamefont
  {Hsieh}}, \bibinfo {author} {\bibfnamefont {D.}~\bibnamefont {Qian}},
  \bibinfo {author} {\bibfnamefont {L.}~\bibnamefont {Wray}}, \bibinfo {author}
  {\bibfnamefont {Y.}~\bibnamefont {Xia}}, \bibinfo {author} {\bibfnamefont
  {Y.~S.}\ \bibnamefont {Hor}}, \bibinfo {author} {\bibfnamefont {R.~J.}\
  \bibnamefont {Cava}}, \ and\ \bibinfo {author} {\bibfnamefont {M.~Z.}\
  \bibnamefont {Hasan}},\ }\bibfield  {title} {\enquote {\bibinfo {title} {A
  topological dirac insulator in a quantum spin hall phase},}\ }\href {\doibase
  10.1038/nature06843} {\bibfield  {journal} {\bibinfo  {journal} {Nature}\
  }\textbf {\bibinfo {volume} {452}},\ \bibinfo {pages} {970--974} (\bibinfo
  {year} {2008})}\BibitemShut {NoStop}%
\bibitem [{\citenamefont {Sato}\ \emph {et~al.}(2011)\citenamefont {Sato},
  \citenamefont {Segawa}, \citenamefont {Kosaka}, \citenamefont {Souma},
  \citenamefont {Nakayama}, \citenamefont {Eto}, \citenamefont {Minami},
  \citenamefont {Ando},\ and\ \citenamefont {Takahashi}}]{sato-nat11}%
  \BibitemOpen
  \bibfield  {author} {\bibinfo {author} {\bibfnamefont {T.}~\bibnamefont
  {Sato}}, \bibinfo {author} {\bibfnamefont {Kouji}\ \bibnamefont {Segawa}},
  \bibinfo {author} {\bibfnamefont {K.}~\bibnamefont {Kosaka}}, \bibinfo
  {author} {\bibfnamefont {S.}~\bibnamefont {Souma}}, \bibinfo {author}
  {\bibfnamefont {K.}~\bibnamefont {Nakayama}}, \bibinfo {author}
  {\bibfnamefont {K.}~\bibnamefont {Eto}}, \bibinfo {author} {\bibfnamefont
  {T.}~\bibnamefont {Minami}}, \bibinfo {author} {\bibfnamefont {Yoichi}\
  \bibnamefont {Ando}}, \ and\ \bibinfo {author} {\bibfnamefont
  {T.}~\bibnamefont {Takahashi}},\ }\bibfield  {title} {\enquote {\bibinfo
  {title} {Unexpected mass acquisition of dirac fermions at the quantum phase
  transition of a topological insulator},}\ }\href {\doibase 10.1038/nphys2058}
  {\bibfield  {journal} {\bibinfo  {journal} {Nature Physics}\ }\textbf
  {\bibinfo {volume} {7}},\ \bibinfo {pages} {840--844} (\bibinfo {year}
  {2011})}\BibitemShut {NoStop}%
\bibitem [{\citenamefont {Li}\ \emph {et~al.}(2021)\citenamefont {Li},
  \citenamefont {Wang}, \citenamefont {Mao}, \citenamefont {Ma}, \citenamefont
  {Huang}, \citenamefont {Dai},\ and\ \citenamefont {Niu}}]{runhan21}%
  \BibitemOpen
  \bibfield  {author} {\bibinfo {author} {\bibfnamefont {Runhan}\ \bibnamefont
  {Li}}, \bibinfo {author} {\bibfnamefont {Hao}\ \bibnamefont {Wang}}, \bibinfo
  {author} {\bibfnamefont {Ning}\ \bibnamefont {Mao}}, \bibinfo {author}
  {\bibfnamefont {Hongkai}\ \bibnamefont {Ma}}, \bibinfo {author}
  {\bibfnamefont {Baibiao}\ \bibnamefont {Huang}}, \bibinfo {author}
  {\bibfnamefont {Ying}\ \bibnamefont {Dai}}, \ and\ \bibinfo {author}
  {\bibfnamefont {Chengwang}\ \bibnamefont {Niu}},\ }\bibfield  {title}
  {\enquote {\bibinfo {title} {Engineering antiferromagnetic topological
  insulator by strain in two-dimensional rare-earth pnictide
  {EuCd$_2$Sb$_2$}},}\ }\href {\doibase 10.1063/5.0063353} {\bibfield
  {journal} {\bibinfo  {journal} {Applied Physics Letters}\ }\textbf {\bibinfo
  {volume} {119}},\ \bibinfo {pages} {173105} (\bibinfo {year} {2021})},\
  \Eprint {http://arxiv.org/abs/https://doi.org/10.1063/5.0063353}
  {https://doi.org/10.1063/5.0063353} \BibitemShut {NoStop}%
\bibitem [{\citenamefont {Hohenberg}\ and\ \citenamefont
  {Kohn}(1964)}]{hohenberg-pr64}%
  \BibitemOpen
  \bibfield  {author} {\bibinfo {author} {\bibfnamefont {P.}~\bibnamefont
  {Hohenberg}}\ and\ \bibinfo {author} {\bibfnamefont {W.}~\bibnamefont
  {Kohn}},\ }\bibfield  {title} {\enquote {\bibinfo {title} {Inhomogeneous
  electron gas},}\ }\href {\doibase 10.1103/PhysRev.136.B864} {\bibfield
  {journal} {\bibinfo  {journal} {Phys. Rev.}\ }\textbf {\bibinfo {volume}
  {136}},\ \bibinfo {pages} {B864--B871} (\bibinfo {year} {1964})}\BibitemShut
  {NoStop}%
\bibitem [{\citenamefont {Kohn}\ and\ \citenamefont {Sham}(1965)}]{kohn-pr65}%
  \BibitemOpen
  \bibfield  {author} {\bibinfo {author} {\bibfnamefont {W.}~\bibnamefont
  {Kohn}}\ and\ \bibinfo {author} {\bibfnamefont {L.~J.}\ \bibnamefont
  {Sham}},\ }\bibfield  {title} {\enquote {\bibinfo {title} {Self-consistent
  equations including exchange and correlation effects},}\ }\href {\doibase
  10.1103/PhysRev.140.A1133} {\bibfield  {journal} {\bibinfo  {journal} {Phys.
  Rev.}\ }\textbf {\bibinfo {volume} {140}},\ \bibinfo {pages} {A1133--A1138}
  (\bibinfo {year} {1965})}\BibitemShut {NoStop}%
\bibitem [{\citenamefont {Kresse}\ and\ \citenamefont
  {Furthm\"uller}(1996)}]{kresse-prb96}%
  \BibitemOpen
  \bibfield  {author} {\bibinfo {author} {\bibfnamefont {G.}~\bibnamefont
  {Kresse}}\ and\ \bibinfo {author} {\bibfnamefont {J.}~\bibnamefont
  {Furthm\"uller}},\ }\bibfield  {title} {\enquote {\bibinfo {title} {Efficient
  iterative schemes for ab initio total-energy calculations using a plane-wave
  basis set},}\ }\href {\doibase 10.1103/PhysRevB.54.11169} {\bibfield
  {journal} {\bibinfo  {journal} {Phys. Rev. B}\ }\textbf {\bibinfo {volume}
  {54}},\ \bibinfo {pages} {11169--11186} (\bibinfo {year} {1996})}\BibitemShut
  {NoStop}%
\bibitem [{\citenamefont {Kresse}\ and\ \citenamefont
  {Joubert}(1999)}]{kresse-prb99}%
  \BibitemOpen
  \bibfield  {author} {\bibinfo {author} {\bibfnamefont {G.}~\bibnamefont
  {Kresse}}\ and\ \bibinfo {author} {\bibfnamefont {D.}~\bibnamefont
  {Joubert}},\ }\bibfield  {title} {\enquote {\bibinfo {title} {From ultrasoft
  pseudopotentials to the projector augmented-wave method},}\ }\href {\doibase
  10.1103/PhysRevB.59.1758} {\bibfield  {journal} {\bibinfo  {journal} {Phys.
  Rev. B}\ }\textbf {\bibinfo {volume} {59}},\ \bibinfo {pages} {1758--1775}
  (\bibinfo {year} {1999})}\BibitemShut {NoStop}%
\bibitem [{\citenamefont {Perdew}\ \emph {et~al.}(1996)\citenamefont {Perdew},
  \citenamefont {Burke},\ and\ \citenamefont {Ernzerhof}}]{perdew-prl96}%
  \BibitemOpen
  \bibfield  {author} {\bibinfo {author} {\bibfnamefont {John~P.}\ \bibnamefont
  {Perdew}}, \bibinfo {author} {\bibfnamefont {Kieron}\ \bibnamefont {Burke}},
  \ and\ \bibinfo {author} {\bibfnamefont {Matthias}\ \bibnamefont
  {Ernzerhof}},\ }\bibfield  {title} {\enquote {\bibinfo {title} {Generalized
  gradient approximation made simple},}\ }\href {\doibase
  10.1103/PhysRevLett.77.3865} {\bibfield  {journal} {\bibinfo  {journal}
  {Phys. Rev. Lett.}\ }\textbf {\bibinfo {volume} {77}},\ \bibinfo {pages}
  {3865--3868} (\bibinfo {year} {1996})}\BibitemShut {NoStop}%
\bibitem [{\citenamefont {Togo}\ and\ \citenamefont {Tanaka}(2015)}]{phonopy}%
  \BibitemOpen
  \bibfield  {author} {\bibinfo {author} {\bibfnamefont {Atsushi}\ \bibnamefont
  {Togo}}\ and\ \bibinfo {author} {\bibfnamefont {Isao}\ \bibnamefont
  {Tanaka}},\ }\bibfield  {title} {\enquote {\bibinfo {title} {First principles
  phonon calculations in materials science},}\ }\href {\doibase
  https://doi.org/10.1016/j.scriptamat.2015.07.021} {\bibfield  {journal}
  {\bibinfo  {journal} {Scripta Materialia}\ }\textbf {\bibinfo {volume}
  {108}},\ \bibinfo {pages} {1--5} (\bibinfo {year} {2015})}\BibitemShut
  {NoStop}%
\bibitem [{\citenamefont {Gao}\ \emph {et~al.}(2021)\citenamefont {Gao},
  \citenamefont {Wu}, \citenamefont {Persson},\ and\ \citenamefont
  {Wang}}]{gao-cpc21}%
  \BibitemOpen
  \bibfield  {author} {\bibinfo {author} {\bibfnamefont {Jiacheng}\
  \bibnamefont {Gao}}, \bibinfo {author} {\bibfnamefont {Quansheng}\
  \bibnamefont {Wu}}, \bibinfo {author} {\bibfnamefont {Clas}\ \bibnamefont
  {Persson}}, \ and\ \bibinfo {author} {\bibfnamefont {Zhijun}\ \bibnamefont
  {Wang}},\ }\bibfield  {title} {\enquote {\bibinfo {title} {Irvsp: To obtain
  irreducible representations of electronic states in the vasp},}\ }\href
  {\doibase https://doi.org/10.1016/j.cpc.2020.107760} {\bibfield  {journal}
  {\bibinfo  {journal} {Computer Physics Communications}\ }\textbf {\bibinfo
  {volume} {261}},\ \bibinfo {pages} {107760} (\bibinfo {year}
  {2021})}\BibitemShut {NoStop}%
\bibitem [{\citenamefont {Herath}\ \emph {et~al.}(2020)\citenamefont {Herath},
  \citenamefont {Tavadze}, \citenamefont {He}, \citenamefont {Bousquet},
  \citenamefont {Singh}, \citenamefont {Muñoz},\ and\ \citenamefont
  {Romero}}]{pyprocar}%
  \BibitemOpen
  \bibfield  {author} {\bibinfo {author} {\bibfnamefont {Uthpala}\ \bibnamefont
  {Herath}}, \bibinfo {author} {\bibfnamefont {Pedram}\ \bibnamefont
  {Tavadze}}, \bibinfo {author} {\bibfnamefont {Xu}~\bibnamefont {He}},
  \bibinfo {author} {\bibfnamefont {Eric}\ \bibnamefont {Bousquet}}, \bibinfo
  {author} {\bibfnamefont {Sobhit}\ \bibnamefont {Singh}}, \bibinfo {author}
  {\bibfnamefont {Francisco}\ \bibnamefont {Muñoz}}, \ and\ \bibinfo {author}
  {\bibfnamefont {Aldo~H.}\ \bibnamefont {Romero}},\ }\bibfield  {title}
  {\enquote {\bibinfo {title} {Pyprocar: A python library for electronic
  structure pre/post-processing},}\ }\href {\doibase
  https://doi.org/10.1016/j.cpc.2019.107080} {\bibfield  {journal} {\bibinfo
  {journal} {Computer Physics Communications}\ }\textbf {\bibinfo {volume}
  {251}},\ \bibinfo {pages} {107080} (\bibinfo {year} {2020})}\BibitemShut
  {NoStop}%
\end{thebibliography}%

\end{document}

%===========================%
% TITLE PAGE                %
%===========================%

\title{ Supplementary Information for ``Engineering magnetic topological insulators in Eu$_5$\textit{M$_2$X$_6$} Zintls"}

\author{Nicodemos Varnava}
\thanks{Corresponding author} 
\email{nvarnava@physics.rutgers.edu}
\affiliation{
Department of Physics \& Astronomy, Rutgers University,
Piscataway, New Jersey 08854, USA}

\author{Tanya Berry}
\affiliation{Institute for Quantum Matter and William H. Miller III Department of Physics and Astronomy, The Johns Hopkins University, Baltimore, Maryland 21218, USA}
\affiliation{Department of Chemistry, The Johns Hopkins University, Baltimore, Maryland 21218, USA}
\affiliation{Department of Materials Science and Engineering, The Johns Hopkins University, Baltimore, Maryland 21218, United States
}

\author{Tyrel M. McQueen}
\affiliation{Institute for Quantum Matter and William H. Miller III Department of Physics and Astronomy, The Johns Hopkins University, Baltimore, Maryland 21218, USA}
\affiliation{Department of Chemistry, The Johns Hopkins University, Baltimore, Maryland 21218, USA}
\affiliation{Department of Materials Science and Engineering, The Johns Hopkins University, Baltimore, Maryland 21218, United States
}

\author{David Vanderbilt}
\affiliation{
Department of Physics \& Astronomy, Rutgers University,
Piscataway, New Jersey 08854, USA}

{
\let\clearpage\relax
\maketitle
} 

%=================================================
\section{Magnetic and non-magnetic band structure of E\lowercase{u}$_5$I\lowercase{n}$_2$S\lowercase{b}$_6$}\seclabel{eu5in2sb6_bands}

\fref{mag_nomag} shows that band structure of Eu$_5$In$_2$Sb$_6$ when (a) the Eu $f$-electrons are set in the core, (b) when an A-type AFM configuration is assumed. In both case the $\zt$ index is trivial. However, the band gap is sensitive to the magnetic configuration and the A-type AFM configuration appears to be metallic in contrast with the experimental observation that Eu$_5$In$_2$Sb$_6$ is in fact a narrow-gap semiconductor. Note that the presence of magnetism enhances moves the the compound further away from the CET limit. This is true for the compounds obtain from Eu$_5$In$_2$Sb$_6$ by chemical substitution.

\begin{figure*}[h]
\centering\includegraphics[width=7.4in]{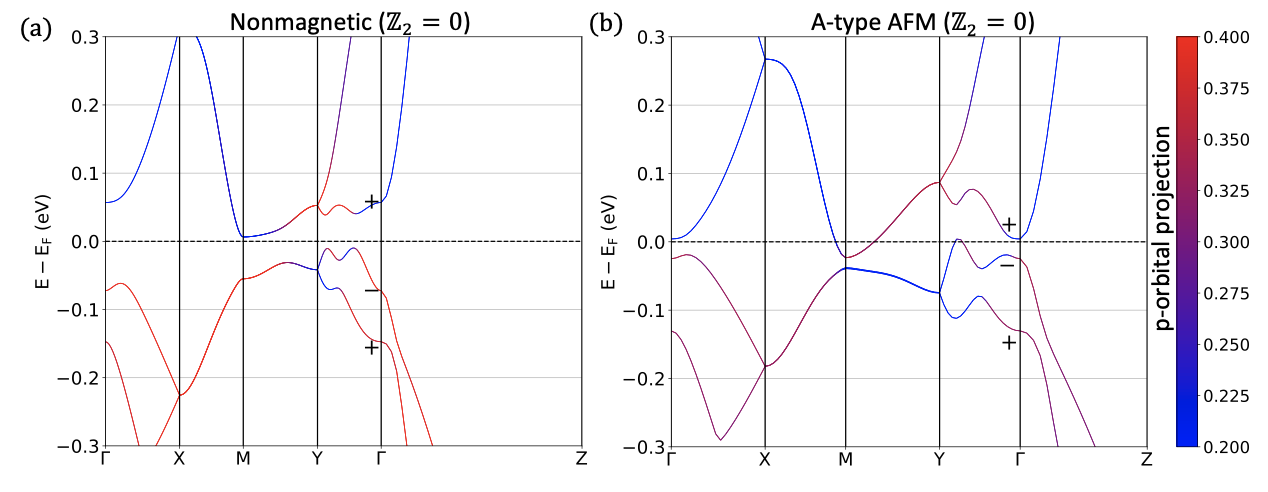}
\caption{(a) Non-magnetic (b) A-type AFM band structure of Eu$_5$In$_2$Sb$_6$.}
\label{fig:mag_nomag}
\end{figure*}

\newpage
\section{Degeneracies and the parity criterion}\seclabel{appendix}

Here we show that only band inversions at $\Gamma$ and Z can alter the axion $\zt$ index for a nonmagnetic material in the centrosymmetric SG 55. As explained below, this is related\cite{chen-prb16,wieder-sc18} to the non-symmorphic nature of SG 55. Briefly, states at all time reversal invariant momenta (TRIM) except $\Gamma$ and Z form 4-dimensional irreducible representations (irreps) for which the number of odd-parity states is fixed modulo four. In view of the Fu-Kane criterion for the strong $\zt$ index\cite{fu-prb07}, band inversions at these TRIM points cannot change the $\zt$ index. Indeed, if $n^{-}_K$ is the total number of odd-parity states at a given TRIM point, then the criterion can be expressed as 
\beq\label{eq:fu-kane}
\prod_{\k\in {\rm TRIM}} (-1)^{n^{-}_\k/2} \in \zt,
\eeq
which remains unaffected when $n^{-}_K$ is fixed modulo four.

We first explain how the 4-fold degeneracies arise. SG 55 can be regarded as generated by two glide mirrors $g_x=\{m_x|1/2,1/2,0\}$ and $g_y=\{m_y|1/2,1/2,0\}$ together with inversion $I$, all of which are good symmetries at all eight of the TRIM (see Table I).  We can use the fact that $\{g_x,g_y\}=0$ to show that if $\ket{\psi_\k}$ is an eigenstate of $g_x$, i.e., $g_x\ket{\psi_\k} =\alpha\ket{\psi_\k}$,  then  $g_x(g_y\ket{\psi_\k}) =-\alpha(g_y\ket{\psi_\k})$. This implies $g_y\ket{\psi_\k}$, which has the same energy as $\ket{\psi_\k}$, is also an eigenstate of $g_x$ with opposite eigenvalue. In addition, the Kramer's partners $T\ket{\psi_\k}$ and $Tg_y\ket{\psi_\k}$ will have $g_x$ eigenvalues $\alpha^*$ and $-\alpha^*$ respectively. Taking into account that the $g_x$ eigenvalues are $\alpha=\pm i(\pm 1)$ at $k_y=0(\pi)$, we can conclude that Kramer's partners, which are necessarily orthogonal to each other, have opposite(same) eigenvalues at $k_y=0(\pi)$. This then implies that $\ket{g_y}$, $T\ket{\psi_\k}$, $g_y\ket{\psi_\k}$ and $Tg_y\ket{\psi_\k}$ are mutually orthogonal at TRIM points where $k_y=\pi$ and they form 4D representations but is inconclusive when $k_y=0$. Finally, note that we can apply the same line of thought after we swap the role of $g_x$ and $g_y$  leading to the conclusion that $\ket{\psi_\k}$, $T\ket{\psi_\k}$, $g_x\ket{\psi_\k}$ and $Tg_x\ket{\psi_\k}$ form 4D dimensional representations at TRIM points where $k_x=\pi$. Putting these together we conclude that TRIM points where either $k_x=\pi$ or $k_y=\pi$ admit representations of at least of degree four. In fact, using the computational tool BANDREP \cite{bradlyn-nat17,vergniory-pre17,elcoro-jac17} of the Bilbao Crystallographic Server\cite{aroyo-acs06,aroyo-zfk06,aroyo-bcc11}, we find that SG 55 admits only 4-dimensional irreps at these TRIM points.

\begin{table}[h]
\label{table1}
\caption{ Symmetry properties of glide mirrors $g_x$, $g_y$ and inversion in SG 55.}
\begin{tabular}{p{4.0cm}p{3.0cm}p{3.0cm}p{3.0cm}}
\hline\hline
Group Symmetry & $g_x$ & $g_y$ & $I$  \\ 
Vector representation &  $(\bar{x}+\frac{1}{2},y+\frac{1}{2},z)$  &  $(\bar{x}+\frac{1}{2},y+\frac{1}{2},z)$ &  $(\bar{x},\bar{y},\bar{z})$ \\ 
Spinor representation &  $-i\sigma_x$  &  $-i\sigma_y$ &  $\sigma_0$ \\ 
Eigenvalues  &  $\alpha = \pm i e^{ik_y/2}$  &  $\beta = \pm i e^{ik_x/2}$ &  $\gamma = \pm 1$ \\ 
\hline\hline 
\end{tabular}
\end{table}

Next we explain how the non-symmorphic symmetries constrain the inversion eigenvalues. Since
%
%\dvm{One reason I don't like the $e^{-ik_x}$ notation is that $k_x=+\pi$ and $k_x=-\pi$ label exactly the same Bloch state. Of course this is consistent with $e^{ik_x}=e^{-ik_x}$ at all of the TRIM, but this also means that the sign in front of $k_x$ is completely meaningless.  If you keep this notation, I would write the two equations below as containing a factor $e^{i(k_x+k_y)}$, since if the signs are meaningless, it is cleaner to use the plus signs.}
%
\beq
\begin{split}
    Ig_x &= g_xIe^{-ik_x+ik_y} \\
    Ig_y &= g_yIe^{ik_x-ik_y}
\end{split}
\eeq
depending on the TRIM point, inversion either commutes or anticommutes with $g_x$ and $g_y$. In either case, $g_x$ and $g_y$ eigenstates are also inversion eigenstates. Considering the set of orthogonal states $\{\ket{g_y},T\ket{\psi_\k},g_y\ket{\psi_\k},Tg_y\ket{\psi_\k}\}$ at the TRIM points with $k_y=\pi$, we observe that Kramer's partners share the same inversion eigenvalues, since they are real numbers, while states related by $g_y$ will have opposite inversion eigenvalues when $k_x + k_y = \pi$, i.e., at the X,Y,U,T points, and same eigenvalues when $k_x + k_y = 2\pi$, i.e., at the S,R points. In the former case, the number of odd-parity states in each irreducible representation is $2$ while in the latter is either $0$ or $4$. In either case band inversions can only change the parity count by $0$ or $4$ which does not affect the $\zt$ index.

\newpage
\section{Displacing the I\lowercase{n} atom in E\lowercase{u}$_5$I\lowercase{n}$_2$S\lowercase{b}$_6$ }\seclabel{displaced}

As we have seen in the main text displacing the In atom by modifying the $y$ component of its Wyckoff position has the geometric effect of distorting the In-Sb tetrahedra by shortening the two In-Sb$_1$ bond and lengthening the In-Sb$_2$ and In-Sb$_3$ bonds. 

The effect of the displacement on the band structure is shown in \fref{displacement}. Here we focus on $k_z=0$ plane, since $k_z=\pi$ remains gapped for all values of displacement, and the lowest conduction and highest valence bands, since they determined the topology and band gaps. For the actual structure, we see that the highest valence bands \fref{displacement}(e) have primarily Sb character while the lowest conduction bands have primarily Eu character. This is what we should expect for a trivial insulator that follows the Zintl concept in which case we are at the limit of complete electron transfer for the Eu$^{+10}$ cations to the [In$_2$Sb$_6$]$^{-10}$ polyanion.

Now as we progressively displace the In atoms, the character of the lowest conduction bands changes from Eu to In signaling that we are moving away from the limit of complete electron transfer. In this way, the energy of the lowest conduction bands decreases and that of the highest valence bands increases. The increased overlap causes a band inversion at $\Gamma$ when SOC is included which makes the $\zt$ index non-trivial.

\begin{figure*}[h]
\centering\includegraphics[width=7.4in]{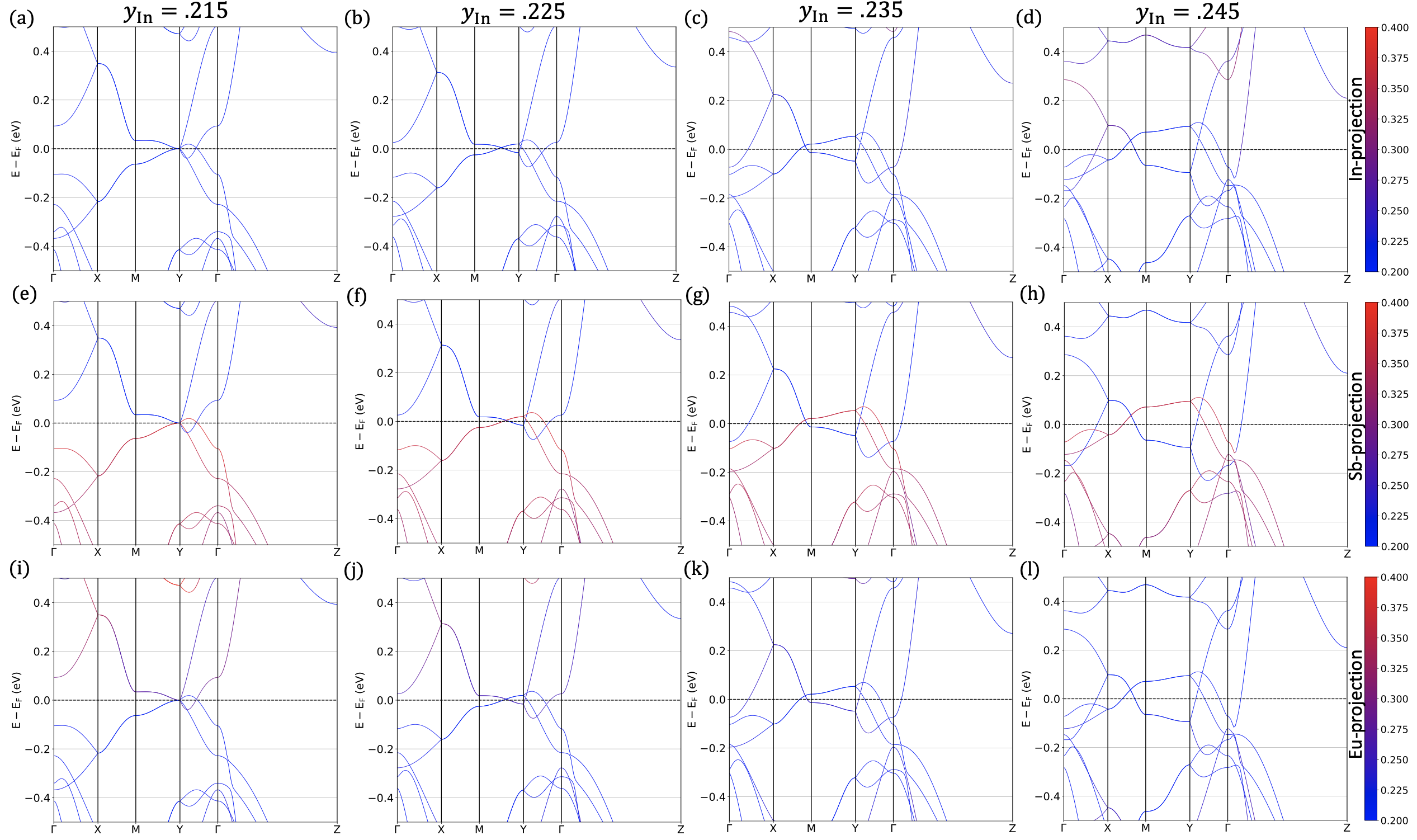}
\caption{Band structures without SOC for different values of the $y$-component of the In atom’s Wyckoff position. From the left to right the band structures interpolate between the actual $y_\text{In}=.2149$ and the one that was reported in \citet{rosa-npjqm2020} $y_\text{In}=2419$. The three rows, correspond to projections on the In (a)-(d), Sb (e)-(h) and Eu (i)-(l) atoms respectively.}
\label{fig:displacement}
\end{figure*}

\newpage
\section{Changing the topology of E\lowercase{u}$_5$I\lowercase{n}$_2$S\lowercase{b}$_6$ through the application of strain}\seclabel{changing_topo}

Here we show that applying compressive uniaxial strain along $c$ or expansive epitaxial strain in the $a-b$ plane has a similar effect as displacing the In atom. In particular, \fref{changing_topo} shows the overlap between conduction and valence bands increases when either of these structural perturbations is applied, indicating that the compound is moving away from the CET limit. For extreme values of, around $\sim10\%$ compressive uniaxial strain or $\sim10\%$ expansive epitaxial strain, the $\zt$ index changes. However, this not only impractical to realize due to the large magnitude of strain but the increased bandwidth of the strained bands results in metallic states.

\begin{figure*}[h]
\centering\includegraphics[width=7.4in]{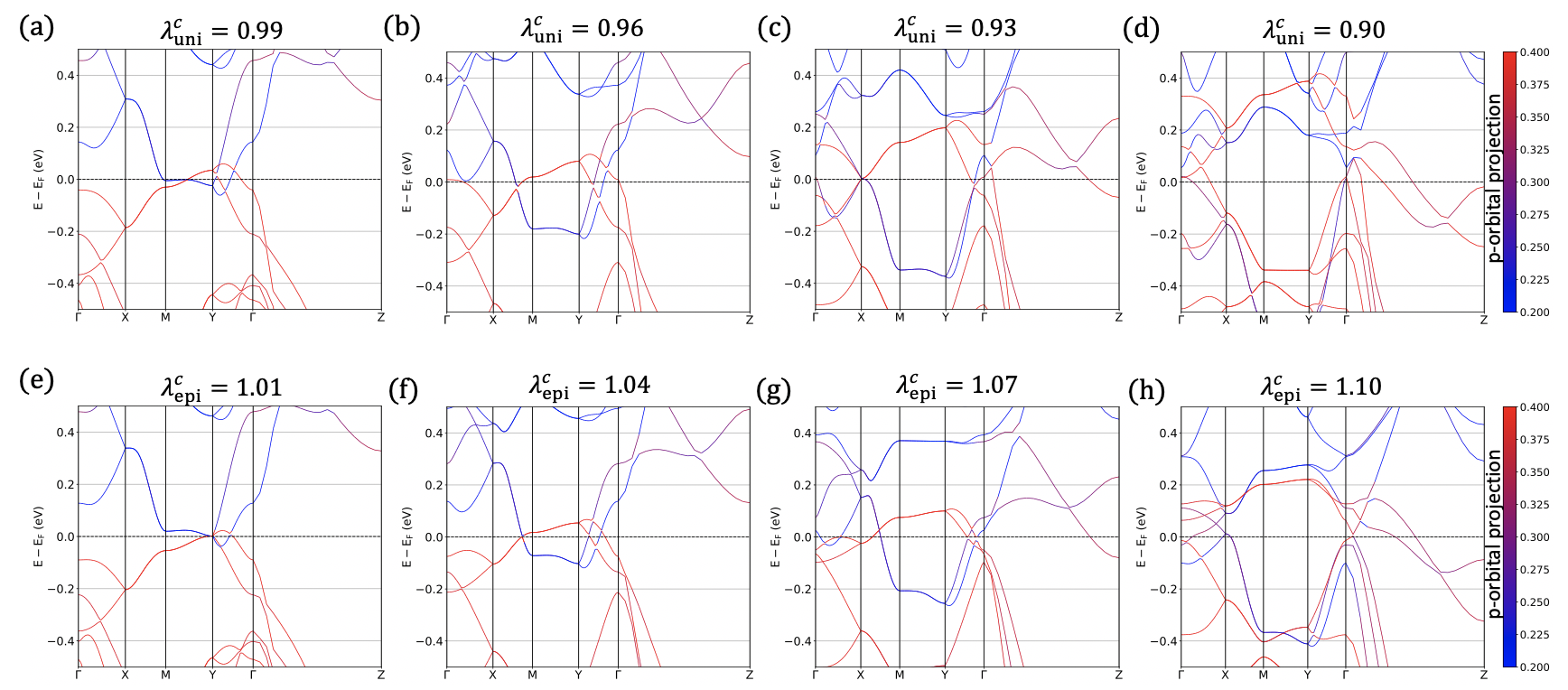}
\caption{Band structures without SOC for different values of uniaxial (a)-(d) and epitaxial (e)-(h) strain. The color mapping indicates the $p$-orbital projection of the states.}
\label{fig:changing_topo}
\end{figure*}

\newpage
\section{Stable G\lowercase{a} and T\lowercase{l} substituted structures}\seclabel{stable}

Starting from  Eu$_5$In$_2$Sb$_6$ we substitute In with Ga or Tl and perform relaxation of the substituted structures (both the internal coordinates and lattice constants). We find that the system remains in SG 55. The structural parameters for Eu$_5$Ga$_2$Sb$_6$ and Eu$_5$Tl$_2$Sb$_6$ are given in Supplementary Tables I,II. To see whether the structures are dynamically stable, we calculate the phonon spectrum, \fref{phonons}. Since the spectrum are positive this indicates that the structures are stable and could be synthesized.

\begin{table}[h]
\label{table:ga}
\centering
\caption{Structure parameters of Eu$_5$Ga$_2$Sb$_6$}
\begin{tabular}{p{1.0cm}p{1.5cm}p{1.5cm}p{1.5cm}p{1.0cm}p{1.0cm}}
\hline\hline 
Atom & x & y & z & Site & Sym.  \\
\hline
Eu$_1$ & 0.33118  &  0.02082  &  0.00000  &  4g  &  ..m \\
Eu$_2$ & 0.00000  &  0.00000  &  0.00000  &  2a  &  ..2/m \\
Eu$_3$ & 0.91044  &  0.75223  &  0.00000  &  4g  &    ..m \\
Sb$_1$ & 0.16526  &  0.82065  &  0.00000  &  4g  &    ..m \\
Sb$_2$ & 0.15615  &  0.09174  &  0.50000  &  4h  &    ..m \\
Sb$_3$ & 0.51984  &  0.09902  &  0.50000  &  4h  &    ..m \\
Ga$_1$ & 0.32645  &  0.21143  &  0.50000  &  4h  &    ..m 
\end{tabular}
\end{table}

\begin{table}[h]
\centering\label{table:tl}
\caption{Structure parameters of Eu$_5$Tl$_2$Sb$_6$}
\begin{tabular}{p{1.0cm}p{1.5cm}p{1.5cm}p{1.5cm}p{1.0cm}p{1.0cm}}
\hline\hline 
Atom & x & y & z & Site & Sym.  \\
\hline
Eu$_1$   &      0.32862  &  0.02364  &  0.00000    &    4g   &    ..m \\
Eu$_2$   &      0.00000  &  0.00000  &  0.00000    &    2a   &  ..2/m \\
Eu$_3$   &      0.91530  &  0.75173  &  0.00000    &    4g   &    ..m \\
Sb$_1$   &      0.16890  &  0.83477  &  0.00000    &    4g   &    ..m \\
Sb$_2$   &      0.14766  &  0.09124  &  0.50000    &    4h   &    ..m \\
Sb$_2$   &      0.53550  &  0.09370  &  0.50000    &    4h   &    ..m \\
Tl$_1$   &      0.33215  &  0.21692  &  0.50000    &    4h   &    ..m
\end{tabular}
\end{table}

\begin{figure*}[h]
\centering\includegraphics[width=5.0in]{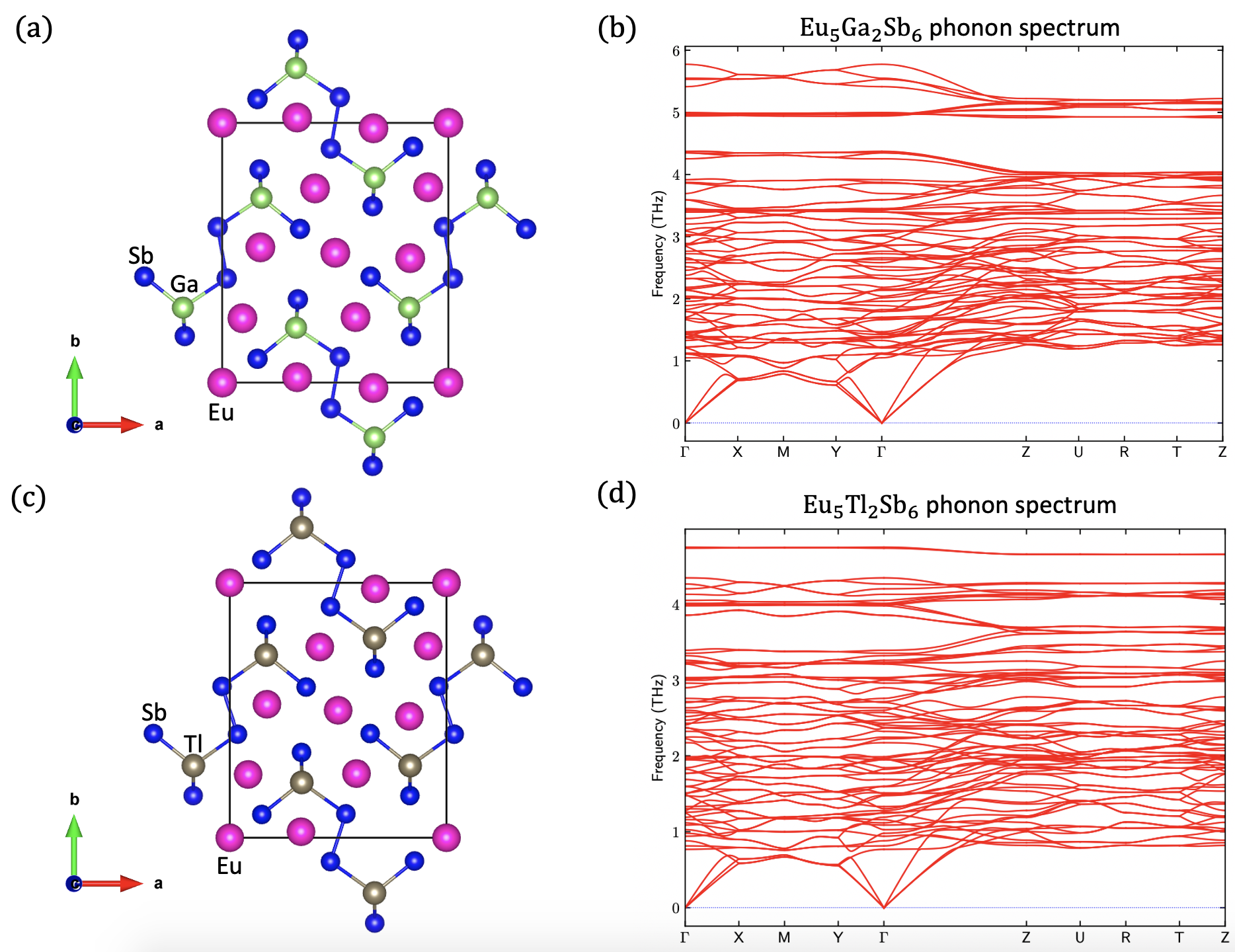}
\caption{(a),(b) Crystal structure and phonon spectrum for Eu$_5$Ga$_2$Sb$_6$ (c),(d) Same for Eu$_5$Tl$_2$Sb$_6$. }
\label{fig:phonons}
\end{figure*}

\section{Band gap engineering using epitaxial strain}\seclabel{uniax_strain_c}

In the main text we showed that uniaxial strain can be used to control the direct and indirect gaps of Eu$_5$In$_2$Bi$_6$. Here we present a similar analysis for epitaxial strain in Eu$_5$In$_2$Bi$_6$.
\fref{eu5in2bi6}(a),(b) show that epitaxial strain with $\lambda^c_\text{epi}<1.00$ or $\lambda^a_\text{epi}>1.00$ enhance the direct and indirect gaps.

\begin{figure*}[h]
\centering\includegraphics[width=7.0in]{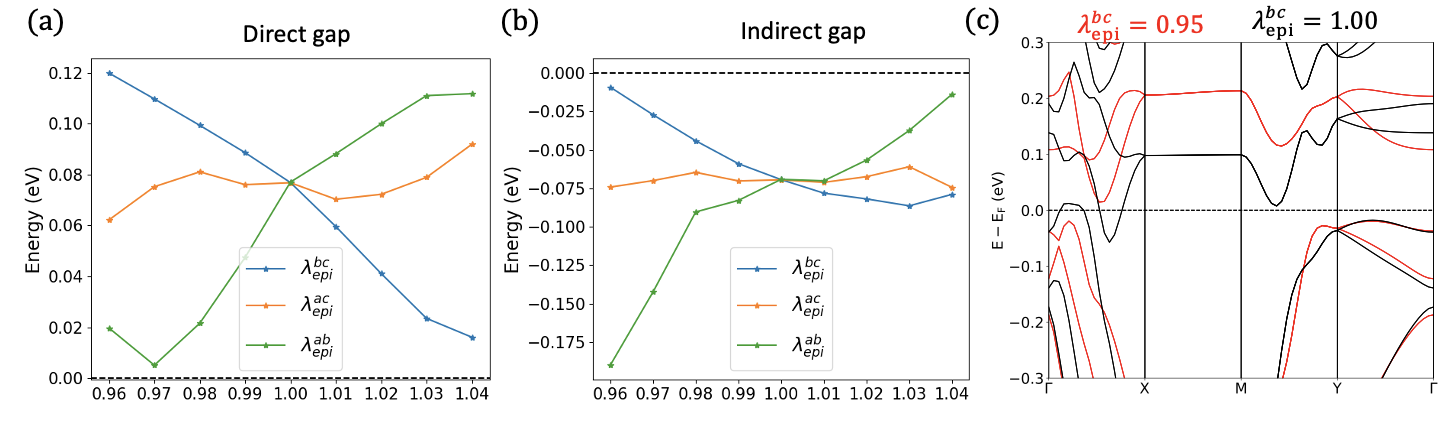}
\caption{Effect of uniaxial strain on (a) direct and  (b) indirect gap. (c) Comparison between band structures with $\lambda^{bc}_\text{uni}=1.00$ and $\lambda^{bc}_\text{uni}=0.95$.}
\label{fig:eu5in2bi6}
\end{figure*}

\bibliographystyle{apsrev4-2}
\bibliography{pap}